\documentclass[runningheads]{llncs}
\usepackage[T1]{fontenc}
\usepackage{graphicx}
\usepackage{adjustbox}
\usepackage{cite}
\usepackage{amsmath,amssymb,amsfonts}
\usepackage{algorithmic}
\usepackage{graphicx}
\usepackage{textcomp}
\usepackage{xcolor}
\usepackage{url}
\usepackage{multirow}
\usepackage{subcaption}
\begin{document}
\title{Augmenting Bankruptcy Prediction using Reported Behavior of Corporate Restructuring
}
\author{Xinlin Wang\inst{1}\orcidID{0000-0003-2275-9424} \and
Mats Brorsson\inst{1}\orcidID{0000-0002-9637-2065} }
\authorrunning{X. Wang and M. Brorsson}
\institute{Interdisciplinary Centre for Security, Reliability and Trust, University of Luxembourg, L-1855 Kirchberg, Luxembourg \\
\email{xinlin.wang@uni.lu, mats.brorsson@uni.lu}}
\maketitle              
\begin{abstract}
Credit risk assessment of a company is commonly conducted by utilizing financial ratios that are derived from its financial statements. However, this approach may not fully encompass other significant aspects of a company. We propose the utilization of a hybrid dataset that combines financial statements with information about corporate restructuring behavior in order to construct diverse machine learning models to predict bankruptcy. Utilizing a hybrid data set provides a more comprehensive and holistic perspective on a company's financial position and the dynamics of its business operations.
The experiments were carried out using publicly available records of all the files submitted by small and medium-sized enterprises to Luxembourg Business Registers.
We conduct a comparative analysis of bankruptcy prediction using six machine learning models. Furthermore, we validate the effectiveness of the hybrid dataset. In addition to the conventional testing set, we deliberately chose the timeframe encompassing the years of the Covid-19 pandemic as an additional testing set in order to evaluate the robustness of the models.
The experimental results demonstrate that the hybrid data set can improve the performance of the model by 4\%-13\% compared to a single source data set. We also identify suitable models for predicting bankruptcy.

\end{abstract}

\section{Introduction}
Small and medium-sized enterprises (SMEs) are of paramount importance in diverse economies. According to the World Bank~\cite{worldbank}, SMEs constitute approximately 90\% of all companies and play a substantial role in generating over 50\% of global employment. SMEs operate predominantly within localized communities, providing employment opportunities to local residents and acting as drivers of economic advancement by fostering competition, innovation, and increased productivity. Moreover, SMEs demonstrate a strong sense of social responsibility and commitment to sustainability, frequently prioritizing community engagement and environmental preservation~\cite{varga2021defining}.
Therefore, helping SMEs to run in good health is of great importance to society and the economy.

Predicting bankruptcy for SMEs can help enterprises to proactively identify potential risks, adapt their business strategies, and improve their overall competitiveness and stability in a timely manner. Furthermore, bankruptcy prediction can facilitate the assignment of credit ratings for SMEs, providing credit endorsements and enabling them to access better financial services. This, in turn, can promote their growth and development~\cite{rao2021systematic}. Finally, bankruptcy prediction can help governments and social organizations identify potential financial crises and implement proactive measures to mitigate the adverse economic and social consequences.

The origins of bankruptcy prediction models can be traced back to the late 1960s, when the Logit model was proposed by Beaver in 1966~\cite{beaver1966financial} and the Z-score model was proposed by Altman in 1968~\cite{altman1968financial}. Both models were formulated using financial ratios and played a significant role during their respective periods, establishing a fundamental framework for subsequent investigations into the prediction of bankruptcy. Financial ratios are accounting-based ratios used to assess the financial health of a company, typically derived from its financial statements~\cite{damodaran1996corporate}. With the development of the financial industry and the field of data science, numerous studies have been conducted on the prediction of studies primarily rely on accounting-based ratios and employ various models to predict~\cite{altman2008value,kim2010ensemble,liang2016financial,hosaka2019bankruptcy,mselmi2017financial,son2019data}. Although several studies have also incorporated various types of data, including market-based variables~\cite{agarwal2008comparing,chandra2009failure,mai2019deep} and macroeconomic indicators~\cite{khoja2019analysis}, studies on input data (or features) had been a largely under explored domain compared to studies on models.

In this study, our goal is to improve the accuracy of bankruptcy prediction models by including data on reported corporate restructuring behavior, in addition to using accounting-based ratios as input variables.
We used a publicly available dataset from Luxembourg Business Registers (LBR)\footnote{\url{https://www.lbr.lu/}} to implement experiments and validate our hypothesis. Registered companies in Luxembourg are required to submit their basic information, business operation files, and financial statements to LBR. We create a hybrid dataset consisting of accounting-based ratios and features related to restructuring behavior. We compare the bankruptcy prediction results of six machine learning models: logistic regression (LR), random forest (RF), lightGBM (LGB), multilayer perceptron (MLP), convolutional neural network (CNN), and long short-term memory (LSTM). We validate the effectiveness of a hybrid dataset and identify suitable models for predicting bankruptcy. We specifically compare the time periods before and after the pandemic as the testing sets to assess the robustness of the models during the special economic period.

This paper makes several contributions to the field of bankruptcy prediction, including the following:
\begin{itemize}

\item We present the first large-sample bankruptcy prediction using corporate restructuring behavior, which, to the best of our knowledge, has not been explored before;

\item We conduct a comparative study of six well-known bankruptcy prediction models using real-world data from Luxembourg Business Registers;

\item We evaluate the performance of the models in response to Covid-19 pandemic period and analyze the drift of prediction models.

\end{itemize}

\section{Related works}
Researchers often prioritize finding ways to improve the effectiveness of the model while overlooking the importance of studying input data. In \cite{kumar2007bankruptcy}'s review, the authors summarized over 60 studies that apply accounting-based ratios, also known as financial ratios, to various models. In ~\ref{relatedworks}, we have provided a list of recent studies that focus on input data and selected representative works that are based on financial ratios. We present these studies by year of publication, categorizing them according to data category, data type, prediction models, evaluation approaches, sample size and publication year. In recent decades, there has been a notable rise in studies that examine different input data. The development of computer technology and data science has made it easier to collect, store, process, and model data.




\begin{table*}[htpb]
    \centering  
    \footnotesize
    \caption{Studies on Bankruptcy Prediction}
\begin{adjustbox}{center}
    \label{relatedworks}
    \begin{tabular}{|l|p{3cm}|p{1.8cm}|p{2.2cm}|p{2.5cm}|p{1.5cm}|p{1cm}|}
    \hline
    Study & Data Category & Data Type & Prediction Models & Evaluation Approaches& Sample Size & Publish Year\\ \hline 
    \cite{altman2008value}& Financial ratios,basic firm information, reported and compliance, operational risk&Numerical data&Altman's Z-score, generic model&AUC, roc curve &3,462,619&2008\\
    \cite{agarwal2008comparing}&Financial ratios, market-based variables&Numerical data&Black and Scholes models, Altman's Z-score&ROC curve, information content tests&15,384&2008\\
    \cite{chandra2009failure}&Financial ratios, market-based variables&Numerical data&MLP, CART, LR, RF, SVM, ensemble, boosting&Accuracy, sensitivity, specificity & 16816 & 2009\\
    \cite{kim2010ensemble}&Financial ratios&Numerical data& MLP, boosting, bagging&Accuracy ratio, AUC &1458&2009\\
    \cite{lin2010role}&Financial ratios, corporate governance indicators& Numerical data& Altman's Z-score, SVM& Type I error, Type II error, average accuracy, brier score& 108 &2010\\ 
    \cite{olson2012comparative}&Financial ratios&Numerical data&DT, LR, MLP, RBFN, SVM& Correct classification rate&1321&2012\\
    \cite{tinoco2013financial} & Accounting, market and macroeconomic data & Numerical data & LR, Altman's Z-score, MLP&AUC, Gini rank coefficient,Kolmogorov–Smirnov & 23,218 & 2013\\        
    \cite{liang2016financial} & Financial ratios, corporate governance indicators& Numerical data& SVM, KNN, NB,CART, MLP&ROC curve&478 & 2016\\
    \cite{hosaka2019bankruptcy}& Financial ratios& Image data & CNN& Identification rates, ROC curve&7520&2019\\
    \cite{tobback2017bankruptcy}& Financial ratios, relational data & Numerical data, graph data & SVM,GNN&AUC&60,000&2017\\
    \cite{mselmi2017financial} & Financial ratios & Numerical data & LR, ANN, SVM, PLS-DA, SVM-PLS& Confusion matrix, accuracy, sensitivity, specificity, AUC& 212 & 2017\\ 
    \cite{khoja2019analysis} & Financial ratios, macroeconomic indicators, industrial factors & Numerical data & MDS&\centering{/}& 165 & 2019\\
    \cite{mai2019deep}&Accounting-based ratio, market-based variables, textual discolures& numerical data, text data& Word embedding, CNN, DNN &Accuracy ratio, AUC& 11,827&2019\\
    \cite{son2019data}&Financial ratios, basic firm information& numerical data, categorical data&LR, RF, XGBoost, LightGBM, ANN&AUC&977,940&2019\\
    \cite{kou2021bankruptcy}& Basic firm informaion, SME network-based variables, transactional data& Numerical data, graph data, categorical data& LDA, LR, SVM, DT, RF, XGB, NN&AUC& 340,531&2021\\
    \cite{huang2022improving}& Textual sentiment & text & SVM, Bayes, KNN, DT, CNN, LSTM&AUC& 10,034 &2022\\
    \hline
    \end{tabular}
    \end{adjustbox}
    \vspace{-0.2cm}
\end{table*}

In recent years, studies have shown the benefits of incorporating diverse input data into bankruptcy prediction models. These studies have expanded beyond traditional data, such as financial ratios and market-based variables, to explore various types of input data. This work~\cite{schalck2021predicting} confirms that financial ratios are predictive indicators of firm failure. The study also suggests that non-financial variables, such as localization and economic conditions, are drivers of SMEs failure. The study~\cite{khoja2019analysis} combines financial ratios and macroeconomic data to analyze their impact on firms, providing evidence for the reliability of macroeconomic data. Another study~\cite{kou2021bankruptcy} focused on using SMEs' transaction data for the prediction of bankruptcy, without relying on accounting data. The results showed that this approach outperformed the benchmark method. Some research pairs also include studies on different types of data. The authors~\cite{tobback2017bankruptcy} used shared directors and managers to establish a connection between two companies and developed a model using relational data to identify the companies with the highest risk. In contrast, this study~\cite{mai2019deep} focuses on using a deep learning model to extract textual information as a complementary variable to accounting and market data to improve prediction accuracy. This paper focuses on corporate restructuring behaviors, such as changes in registered address, management, and corporate regulations. There are two pieces of work~\cite{altman2008value} and ~\cite{liang2016financial} use similar indicators related to corporate restructuring, but they are static and cannot reflect corporate behavior. The present study addresses this gap by focusing on data that reflect changes in corporate behavior.

As demonstrated in table~\ref{relatedworks}, many studies have been dedicated to improving prediction accuracy using various models. Bankruptcy prediction models must be applied practically in the financial industry, necessitating both model accuracy and explainability. The work~\cite{olson2012comparative} compares the accuracy and explainability of different data mining methods for predicting bankruptcy. The study compared algorithms such as neural networks, support vector machines, and decision trees, and concluded that decision trees are both more accurate and easier to interpret compared to neural networks and support vector machines. In ~\cite{lextrait2022scaling} it was demonstrated that LightGBM achieved the highest performance, with fast and cost-effective training, and the model's results could be interpreted using SHAP value analysis. In contrast, authors~\cite{mai2019deep} argue that simple deep learning models outperform other data mining models. This paper will conduct a comparative analysis study by selecting multiple models from ~\ref{relatedworks} which are more universal for data modeling and comparison.
The authors~\cite{grice2001limitations} use the classical bankruptcy prediction models from the study~\cite{zmijewski1984methodological} and the study~\cite{ohlson1980financial} to validate the model performance for different time periods, however, applying these models to a time period other than the one in which they were developed can significantly reduce their accuracy. Furthermore, the study~\cite{papik2023impacts} shows that forecasting models perform significantly worse during crisis periods compared to non-crisis periods. In this paper, we compare the model performance in two periods: pre-Covid19 and post-Covid19, to verify whether the model performance changes due to the pandemic.


In summary, this paper aims to enhance the performance of bankruptcy prediction model by utilizing a hybrid dataset that combines corporate restructuring behavior data with accounting-based ratios. To determine the robustness of the model during the Covid-19 period, separate testing sets will be used for observation.

\section{Methodology}
The main focus of our study is to examine the effectiveness of reported corporate restructuring behavior in predicting bankruptcy. We also aim to analyze the robustness of various bankruptcy models during the Covid-19 pandemic. In this section, we will first present the overall framework for investigating these problems. Then, we will focus on the details of the input data and explain the experimental design.
\subsection{Conceptual framework}

\begin{figure}[h]
  \centering
  \includegraphics[width=0.5\columnwidth]{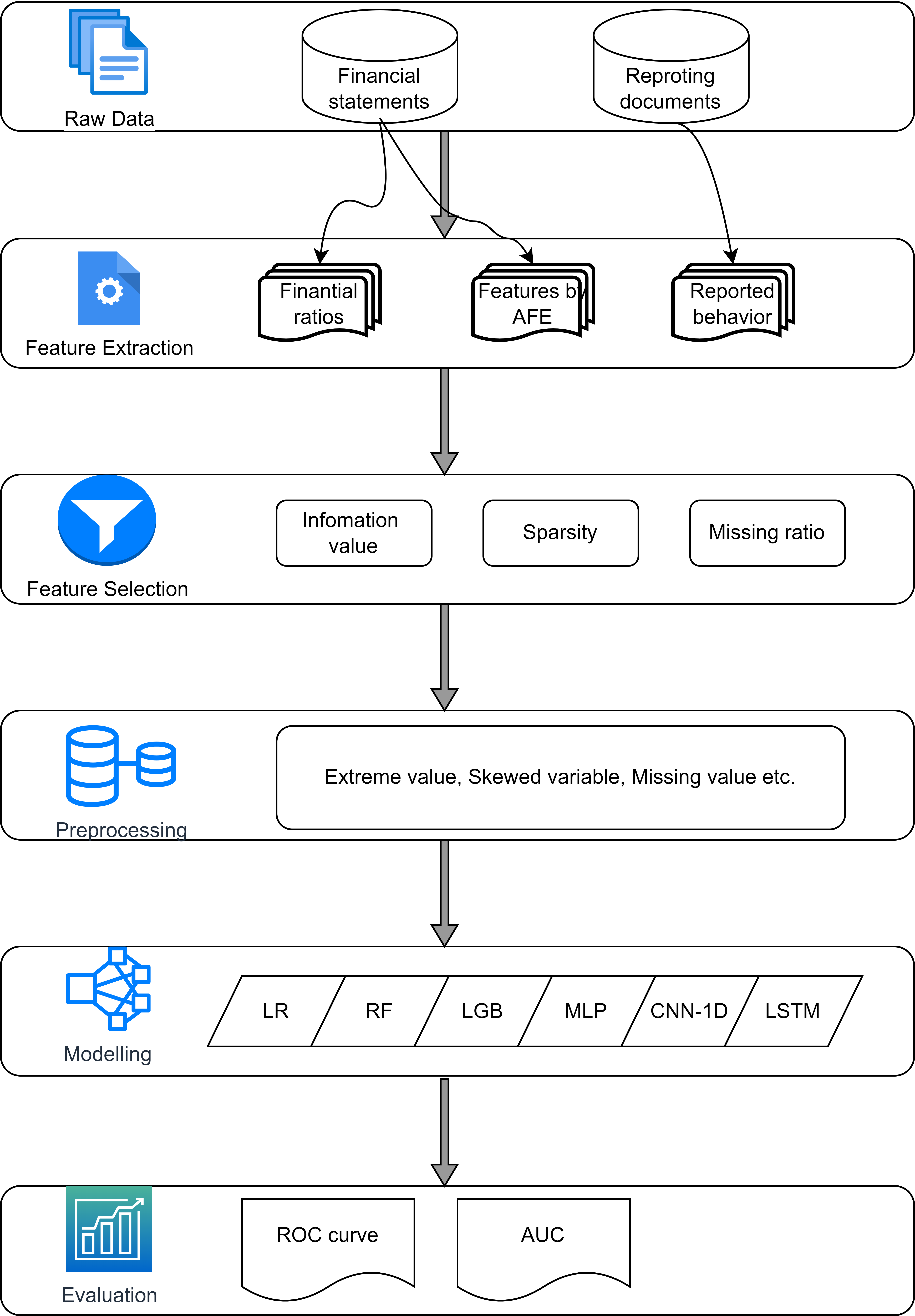}\\
  \caption{Conceptual framework of experimental design }
  \label{pipline}
  \vspace{-0.8cm}
\end{figure}
Fig. \ref{pipline} illustrates the six stages of a framework designed to conduct a comparative study of bankruptcy models using different input data. The data used in the study consists of financial statements and reporting documents. However, since the cash flow statement and profit \& loss statement were not included, the financial statements only consisted of the balance sheet. The reporting documents that companies submit to disclose their operational behaviors are usually classified as textual files.

Different methods are used to extract three types of features from the raw data. The first is financial ratios. Since we do not have a cash flow statement or profit \& loss statement, we will create as many financial ratios as possible. The second type also includes accounting-based features. These variables are constructed using an automatic feature engineering method that we developed in our previous work. This method is capable of creating highly effective features even with limited data~\cite{wang2022financial}. The final type of features is behavior-related features. We design these variables based on corporate restructuring behavior, such as changes in registered addresses, manager resignations, and mergers and acquisitions.

The next stage of the framework involves selecting features from the current variables to eliminate unfavorable and redundant variables caused by sparsity, missing, and repetition. The information value (IV) is an indicator used to measure the predictive power of an independent feature~\cite{howard1966information}. A higher information value indicates that the feature has greater predictive power. The formula for calculating information value is as follows~\cite{siddiqi2012credit}:
\vspace{-0.3cm}
\begin{equation} 
IV = \sum_{i=1}^{n} (\frac{G_i}{G} - \frac{B_i}{B}) * \ln \frac{G_i/G}{B_i/B} 
\vspace{-0.2cm}
\end{equation}
We selected features with an IV value greater than 0.02 and a missing rate less than 0.7.

The data are pre-processed to address missing values, infinite values, and skewed variables, making it more suitable for modeling. We also exclude abnormal samples, such as companies that have submitted financial reports prior to the reference year. We replace infinite values with the highest finite value. In the fifth stage, we assess the effectiveness of behavior-related features by comparing the prediction performance of hybrid datasets that include both behavior-related features and accounting-based ratios with datasets that only contain accounting-based ratios. We trained six popular models, including logistic regression (LR), random forest (RF), LightGBM (LGB), multiple perceptron (MLP), convolutional neural network (CNN) and long-short-term memory (LSTM) to compare their prediction results. We use the receiver operating characteristic curve (ROC curve) and Area under the ROC curve (AUC) as indicators of evaluating the performance of models, which are commonly used and discussed in Section II.

\subsection{Variables \& data}
Since over 90\% of the firms in the Luxembourg industry distribution are finance-related, this paper excludes these firms and focuses only on SMEs as our target samples to avoid an imbalanced distribution of samples.
The state of a company is not static and can change over time, either by being established or going bankrupt. This means that the company may enter or exit the sample set. We utilize a sliding time window approach to sample from the raw data. The sliding-time window is a technique used to extract data from a time series dataset by defining a fixed period of time (window) and moving it forward by a certain interval (step size). This technique allows for continuous monitoring of system states~\cite{zhang2016sliding}.

\begin{figure}[htbp]
    \centering
    \includegraphics[width=0.6\columnwidth]{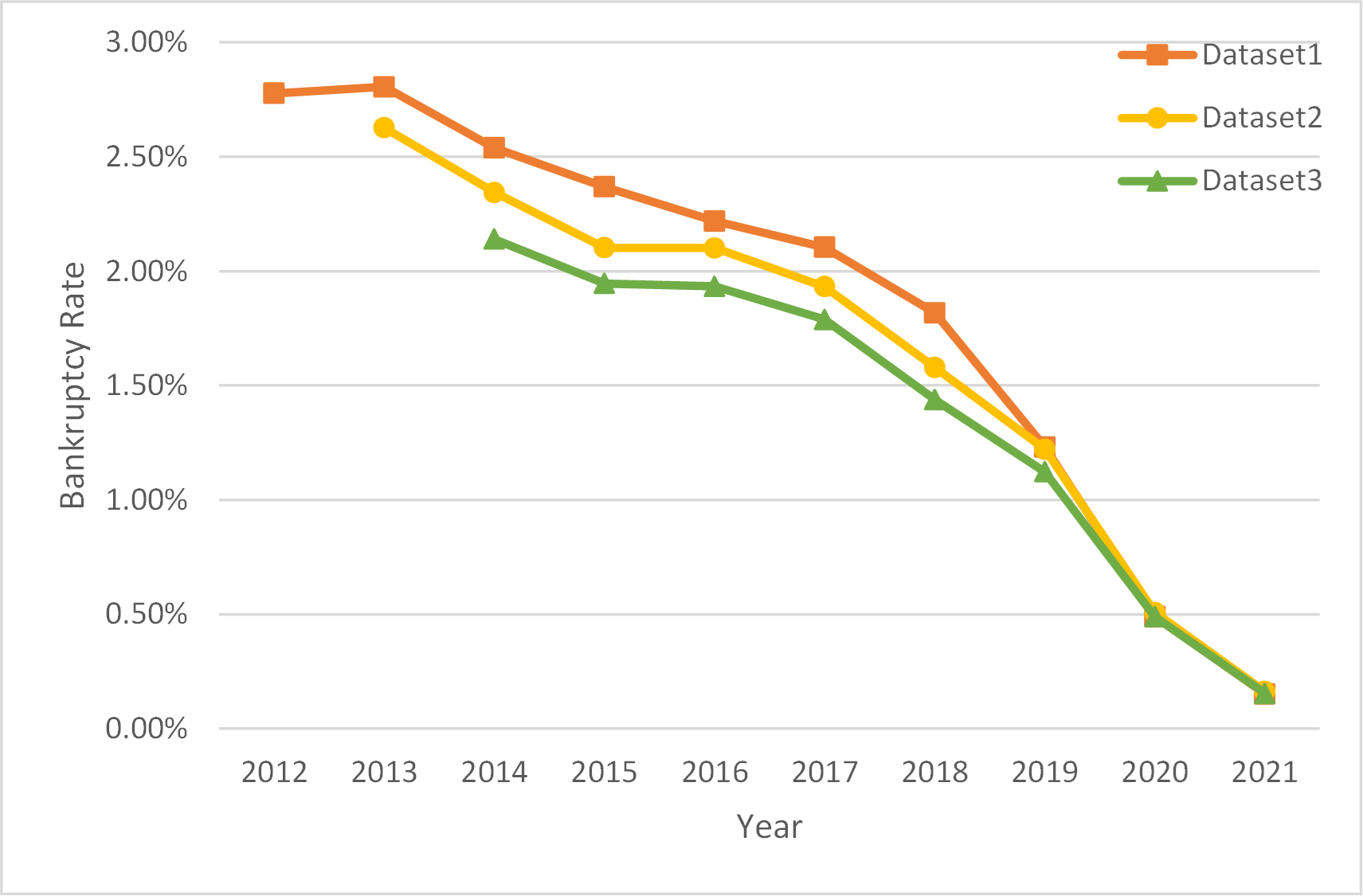}
    \caption{Bankruptcy rate of three datasets from 2012 to 2021 }
    \label{bankruptrate}
    \vspace{-0.8cm}
\end{figure}

\begin{figure}[htbp]
    \vspace{-0.5cm}
    \centering
    \includegraphics[width=0.6\columnwidth]{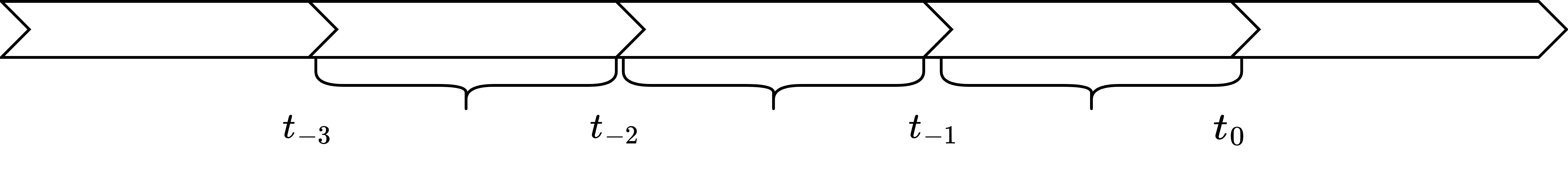}
    \vspace{-0.5cm}
    \caption{Definition of time period}
    \label{timeline}
    \vspace{-0.8cm}
\end{figure}

As of June 2022, there are 74,611 companies in Luxembourg. The average lifespan of companies is approximately 3.5 years. Therefore, we have selected a timeframe of up to 3 years for predicting bankruptcy. We create datasets with three different windows (1-year, 2-year, and 3-year) to predict one step forward (one year). According to the timeline (see Fig.~\ref{timeline}), the three datasets consist of 1-year data from $t_{-1}$ to $t_0$, 2-year data from $t_{-2}$ to $t_0$, and 3-year data from $t_{-3}$ to $t_0$. The sample size of these three datasets, including solvent and bankrupted companies, was summarized in \ref{summary}. However, there is another category of companies with an unknown status. Some companies have not uploaded annual reports or declared bankruptcy, which contributes to the variation in data from year to year. As depicted in the Fig.~\ref{bankruptrate}, the bankruptcy rate of SMEs in Luxembourg has decreased over the past decade. It may indicate that the business conditions of SMEs are improving or that the overall economic environment has improved, resulting in greater stability for SMEs. Additionally, other factors such as policy support or industry changes may also influence the bankruptcy rate of SMEs. It is surprising to find that the bankruptcy rate of SMEs has actually increased during the Covid-19 pandemic, suggesting that fewer SMEs are going bankrupt compared to previous periods. We hypothesize that this could be attributed to government financial assistance during the special period. Some companies may be technically bankrupt but have not yet filed for bankruptcy due to delays in filing.
\begin{table}[htbp]
\vspace{-0.8cm}
    \centering
    \scriptsize{}
    \caption{Summary of three datasets}
    \label{summary}
    \begin{tabular}{|c|c|c|c|c|c|c|}
    \hline
        \multirow{2}{*}{\textbf{Year}}& \multicolumn{2}{|c|}{\textbf{1-year}} & \multicolumn{2}{|c|}{\textbf{2-year}} &\multicolumn{2}{|c|}{\textbf{3-year}}\\ 
        \cline{2-7}
        & Solvent & Bankrupt & Solvent & Bankrupt & Solvent & Bankrupt \\ \hline
        \textbf{2012} & 21738 & 621 & / & / & / & / \\ 
        \textbf{2013} & 23804 & 687 & 17087 & 461 & / & / \\ 
        \textbf{2014} & 25686 & 669 & 18790 & 451 & 16512 & 361 \\ 
        \textbf{2015} & 27331 & 663 & 20301 & 436 & 18188 & 361 \\ 
        \textbf{2016} & 28781 & 653 & 21475 & 461 & 19477 & 384 \\
        \textbf{2017} & 30748 & 661 & 22789 & 449 & 20755 & 378 \\ 
        \textbf{2018} & 32718 & 606 & 24419 & 392 & 22061 & 322 \\
        \textbf{2019} & 34557 & 431 & 25793 & 319 & 23504 & 267 \\
        \textbf{2020} & 36309 & 179 & 27034 & 138 & 24596 & 121 \\
        \textbf{2021} & 22387 & 34 & 17195 & 28 & 15571 & 24 \\ \hline
    \end{tabular}
    \vspace{-0.3cm}
\end{table}

As mentioned earlier, we derive three types of features from raw data: two accounting-based variables and one behavior-based variable. The statistics and descriptions of these features can be found in the table~\ref{variables}. SMEs are not required to prepare and disclose cash flow statements and income statements. Therefore, we can only calculate 18 financial indicators based on the available data~\cite{boguslauskas2011selection,yu2014bankruptcy, zhu2019forecasting, zikeba2016ensemble}. We developed an algorithm for automatic feature engineering ~\cite{wang2022financial} to derive as many useful features as possible from financial statements to address issues caused by the absence of certain financial statements or data quality problems. This algorithm maximizes data mining to generate high-quality features that enhance prediction accuracy. Behavior-based variables are derived from information reported by SMEs regarding corporate restructuring, including both statistical and trend variables.

\begin{table}[h]
\vspace{-0.8cm}
  \centering
  \caption{Description of variables in this study}
  \begin{adjustbox}{center}
  \small
  \begin{tabular}{|p{3cm}|p{12cm}|}
    \hline
    \textbf{Variable} &\textbf{Description}\\
    \hline
    Financial ratios (FR) & current ratio, debt to equity, working capital to total assets ,total liabilities to total assets, equity to total assets, quick ratio ,current assets to total assets ,cash to total assets,cash to current liabilities ,long term debt to equity,total assets growth rate,quick assets to total assets,current assets to current liabilities,(cash or marketable securities) to total assets,total debt to total assets,equity to fixed assets,current assets to total liabilities,short-term liabilities to total assets\\
    \hline
    Automatic feature engineering (AFE)  & automatically generate features from financial statements, which can adapt to any kind of numerical data\\
    \hline
    Reported corporate restructuring behavior-related features (RB) & Modification of name or corporate name, registered office, social object, administrator/manager, daily management delegate, associate, person in charge of checking the accounts, Social capital/social funds, managing director / steering committee, duration, legal form, social exercise, permanent representative of the branch, merger / demerger, depositary, transfer of business assets, assets or business sectors, address, trading name, activities, manager, seat, reason, name, chairman / director, personne autorisée à gérer, administrer et signer, person with the power to commit the company, ministerial approval\\
    \hline
  \end{tabular}
   \label{variables}
   \end{adjustbox}
   \vspace{-0.6cm}
\end{table}


\subsection{Experimental setup}
\subsubsection{Dataset Description}
We divide the datasets into two parts: the training set and the testing set, as outlined in table~\ref{datasets}. To maintain consistency between the training set and the testing set, we divided the data from 2012 to 2018 into a 70\% training set and a 30\% testing set. Additionally, we created two additional testing sets: one using solvent and bankrupt SMEs from 2019 as a pre-Covid testing set, and another using solvent and bankrupt SMEs from 2020 and 2021 as a post-Covid testing set. To train our models, we utilized the 5-fold cross-validation method and did not set aside a separate validation set. Table~\ref{datasets} have a bankruptcy rate below 3\%, making them highly imbalanced. The negative datasets are typically large in size, so we used the under-sampling method during data preprocessing to balance the rate to 25\%.

\begin{table}[htbp]
    \centering
    \caption{Summary of datasets splitting}
    \small
    \begin{tabular}{|l|l|l|l|l|}
    \hline
        \multicolumn{2}{|c|}{\multirow{2}{*}{}}& \textbf{1-year} & \textbf{2-year} & \textbf{3-year} \\ \hline
        \multirow{3}{*}{\textbf{Training}} & Solvent (Negative)  & 110805 & 87467 & 67920  \\ 
        & Bankrupt (Positive)& 2625 & 1797 & 1244 \\
        & Bankruptcy Rate & 2.31\% & 2.01\% & 1.80\% \\ \hline
        \multirow{3}{*}{\textbf{Testing}} & Solvent (Negative) & 47458 & 37403 & 29081 \\ 
        & Bankrupt (Positive)& 1155 & 853 & 562 \\
        & Bankruptcy Rate & 2.38\% & 2.23\% & 1.90\% \\ \hline
        \multirow{3}{*}{\textbf{Pre-Covid}} & Solvent (Negative) & 28730 & 25793 & 23504 \\ 
        & Bankrupt (Positive)& 368 & 319 & 267 \\
        & Bankruptcy Rate & 1.26\% & 1.22\% & 1.12\% \\ \hline
        \multirow{3}{*}{\textbf{Post-Covid}} & Solvent (Negative) & 48846 & 44229 & 40167 \\
        & Bankrupt (Positive) & 181 & 166 & 145 \\
        & Bankruptcy Rate & 0.37\% & 0.37\% & 0.36\% \\ \hline
    \end{tabular}
    \label{datasets}
    \vspace{-0.6cm}
\end{table}

\subsubsection{Models}
In this paper, we have chosen six bankruptcy prediction models, which include statistical, machine learning, and deep learning models, by synthesizing the statistics from previous studies in Part II. We comprehensively evaluate the behavior-based features by comparing the performance of representative models. The table~\ref{machine} displays the environmental information used for model training.

\begin{table}[htbp]
\vspace{-0.5cm}
\small
    \centering
    \caption{Information of training machine}
    \begin{tabular}{|l|l|}
    \hline
        Device name &         Tesla V100-SXM2-32GB \\ \hline
        Linux version & Red Hat 8.5.0-10 \\ \hline
        Python version & 3.8.6 \\ \hline
        Pytorch version &     1.10.1+cu111 \\ \hline
        Cuda version & 11.1 \\ \hline
        Cudnn version & 8005 \\ \hline
        Sklearn version & 1.2.1 \\ \hline
        Number of GPU & 2 \\ \hline
        Number of CPU & 16 \\ \hline
    \end{tabular}
    \label{machine}
    \vspace{-0.8cm}
\end{table}

\begin{figure}
\vspace{-0.5cm}
    \centering
    \includegraphics[width=0.6\columnwidth]{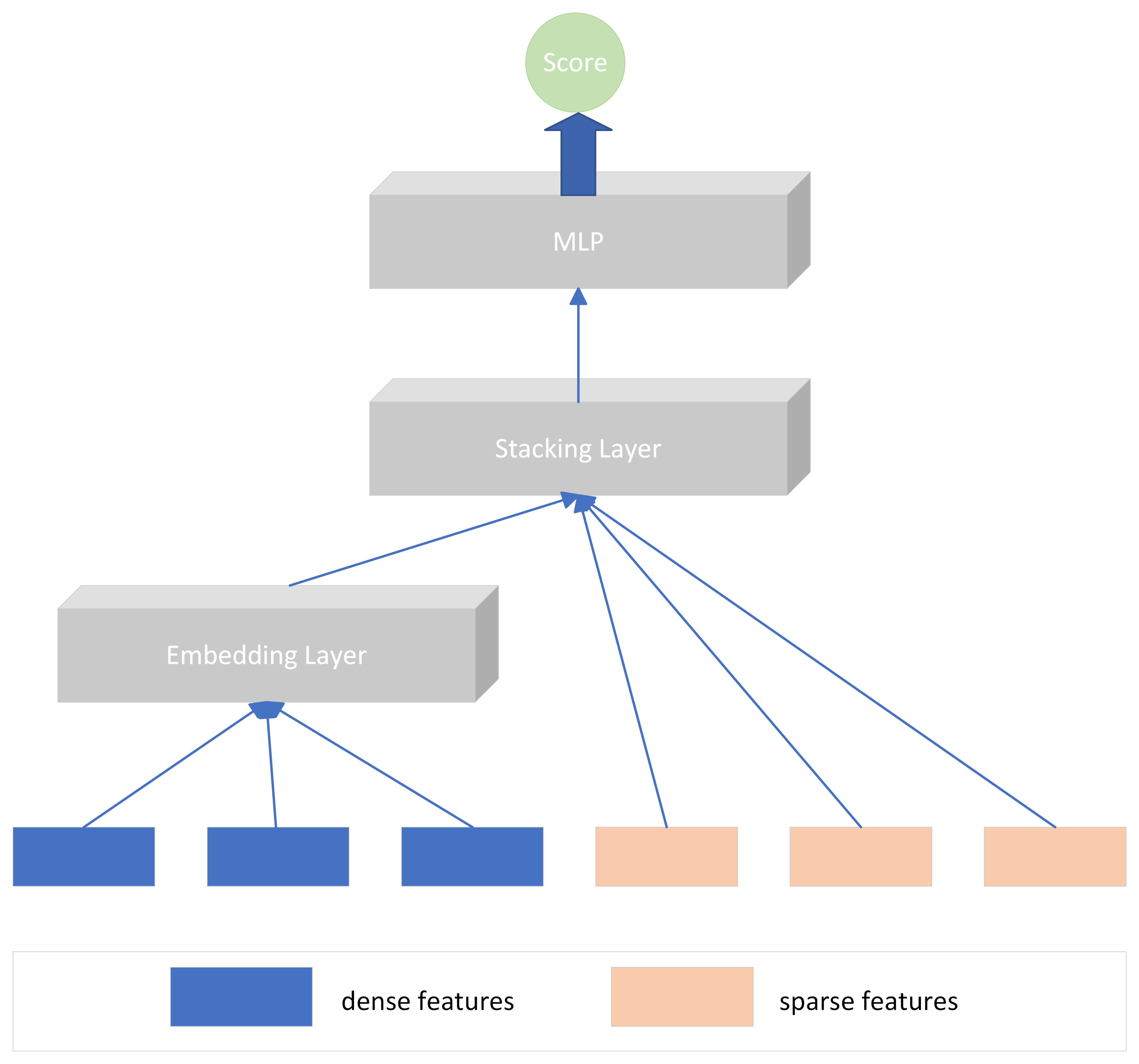}
    \caption{Structure of MLP}
    \label{mlp_structure}
    \vspace{-0.9cm}
\end{figure}

\paragraph{Logistic regression} predicts the likelihood of a binary outcome using one or more predictor variables. Logistic regression models have advantages in bankruptcy prediction due to their simplicity, fast computation, and better results when dealing with smaller datasets. 
In this study, we adopt \textit{LogisticRegression} from \textit{sklearn} package and use \textit{GridSearch} to determine the optimal parameters within a specific range.
\paragraph{Random forest} creates a forest of decision trees, with each tree being trained on a random subset of the data and a random subset of predictor variables. 
Random forest models outperform single decision tree models and other classification models in terms of predictive performance and robustness, and can effectively handle high-dimensional and complex datasets. In this study, we utilize \textit{RandomForestClassifier} from \textit{sklearn} package and employ \textit{GridSearch} to determine the optimal parameters within a specific range.
\paragraph{LightGBM} prioritizes speed and efficiency, specifically for managing large datasets. The method utilizes a gradient-based approach to construct decision trees and incorporates various optimization techniques to accelerate the training process.
In this study, we adopt \textit{LGBMClassifier} from \textit{lightgbm} package and use \textit{GridSearch} to decide the best parameters from a specific range.
\paragraph{Multilayer perceptron} is commonly used for classification and regression tasks. It has the ability to learn complex non-linear relationships between inputs and outputs, making it a powerful tool for various applications. In this study, we incorporate embedding layers for sparse reported behavior features, as depicted in Fig.~\ref{mlp_structure}. We train the model by Pytorch. We choose \textit{BCEWithLogitsLoss} as loss function, \textit{Adam} as optimizer, and \textit{auc} as metric function. We set batch size to 64, epoch to 50 and learning rate to 0.00001.
\paragraph{Convolutional neural network} In this study, we only have tabular data, so we use a one-dimensional CNN (CNN-1D) for prediction. CNN-1D is more effective at capturing local features in the data and has a strong ability to adapt. The convolutional layer extracts features from input data, the pooling layer reduces the number of features and improves model robustness, and the fully connected layer maps the features to the output space for classification. We choose \textit{BCEWithLogitsLoss} as loss function, \textit{Adam} as optimizer, and \textit{auc} as metric function. We set batch size to 1024, epoch equals to 50 and learning rate to 0.00005.
\paragraph{Long short-term memory} aims to address the vanishing gradient problem commonly encountered in traditional recurrent neural networks. The model is capable of retaining long-term dependencies in the input data, making it suitable for various sequence prediction tasks. We reshape the data to fit the time step and features for LSTM in order to predict bankruptcy several years in advance.In this study, we choose \textit{BCEWithLogitsLoss} as loss function, \textit{Adam} as optimizer, and \textit{auc} as metric function. We set batch size to 64, epochs to 50 and learning rate to 0.00001.

\subsubsection{Performance Evaluation}
In selecting the performance measures for the model, we refer to and synthesize previous studies in Section II and select two metrics, AUC and ROC curve, to assess the effectiveness of the model.
\paragraph{Area under the Receiver Operating Characteristic Curve (AUC)} is a performance metric that assesses a classification model's ability to differentiate between positive and negative samples. AUC is not affected by sample imbalance or threshold selection, making it a more comprehensive measure of classifier performance compared to accuracy. The interpretation is straightforward as it summarizes the model's performance with a single scalar value. The formula for calculating AUC is:
\begin{equation} 
\vspace{-0.2cm}
\text{AUC} = \int_{0}^{1} \text{TPR}(FPR^{-1}(t))\, dt
\vspace{-0.2cm}
\end{equation}
\paragraph{Receiver Operating Characteristic Curve (ROC curve)} is a graphical representation of the True Positive Rate (TPR) plotted against the False Positive Rate (FPR) at various classification thresholds. TPR represents the proportion of positive samples correctly classified as positive. On the other hand, FPR represents the proportion of negative samples incorrectly classified as positive. ROC curve is a useful tool for visualizing the trade-off between TPR and FPR at various classification thresholds. The curve is created by plotting the TPR against FPR for every possible classification threshold. It offers a visual representation of the model's performance and helps in selecting the right classification threshold, considering the desired balance between TPR and FPR.
The formula for calculating TPR and FPR is:
\begin{equation}
\vspace{-0.5cm}
\text{FPR} = \frac{\text{False Positives}}{\text{False Positives} + \text{True Negatives}} 
\end{equation}
and 
\begin{equation}
\text{FPR} = \frac{\text{False Positives}}{\text{False Positives} + \text{True Negatives}}
\end{equation}
And we plot ROC curve by 
\begin{equation}
\text{ROC curve: } \text{TPR} \text{ vs } \text{FPR}
\vspace{-0.5cm}
\end{equation}

\section{Results \& Discussion}
\subsection{Features evaluation}
\begin{figure}[htbp]
\vspace{-0.8cm}
    \centering
    \includegraphics[width=0.6\columnwidth]{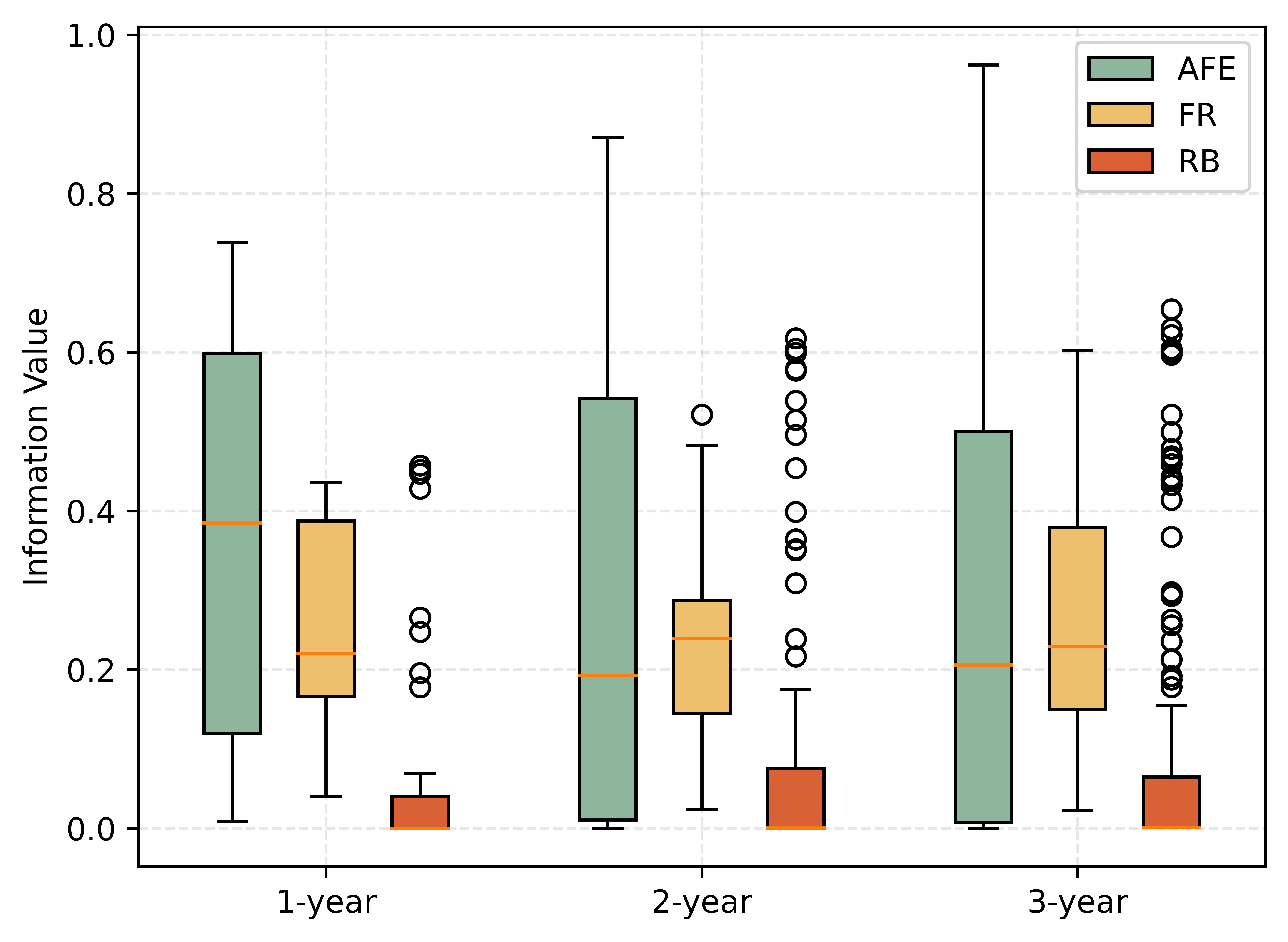}
    \vspace{-0.2cm}
    \caption{IV for features created from AFE, FR and RB}
    \label{iv}
    \vspace{-0.8cm}
\end{figure}

Information value (IV) is a widely used metric for selecting features in binary classification models. Assesses the ability of a feature to predict the target variable by analyzing its relationship. In essence, IV quantifies the amount of information that a feature provides about the target variable. It is commonly used to rank the importance of different features in a predictive model. We calculate the IV for AFE features, financial ratios, and behavior-related features. The results are displayed in Fig.~\ref{iv}. When performing feature selection using IV, features with high IV scores are generally considered more important and informative than those with low IV scores. By eliminating features with low IV scores, we can potentially simplify the model, enhance its performance, and identify the most significant predictors for a specific problem. Additionally, IV provides a standardized and interpretable measure of feature importance that can be easily communicated to stakeholders and decision-makers.

We observe that the number of AFE features is the highest, and most of these features have relatively high IV values. Financial ratios, while fewer in number compared to AFE features, have higher IV values and are less varied. On the other hand, behavior-related features exhibit a wide range of IV values, with some having very high values and the majority clustered towards the lower end of the y-axis. This suggests that these features have little impact on predicting bankruptcy. Behavioral correlation features are often sparse matrices, with the majority of eigenvalues being 0. To mitigate the drawbacks of high coefficient matrices, we will employ feature filtering and summing techniques to maximize the utilization of the data.

\subsection{Ablation experimental results}
We implement the ablation experiments to evaluate if the behaviour-related features can improve the model performance. We create four datasets: AFE, AFE+RB, FR and FR+RB to compare the model performance of with RB features and without RB features. The experiments were carried out on 6 models and 3 time periods. We select 2 out of the 18 results as the representative results and include all the other experimental results in the appendix for reference.
Fig.~\ref{big} summarizes the performance of different features on lightGBM and LSTM by comparing their ROC curves. We use a green line to represent AFE features, a yellow line to represent FR features, a red line to represent AFE and RB features, and a brown line to represent FR and RB features. From this figure, it can be clearly seen that models trained on hybrid datasets of financial and behavior-related features outperform datasets that only include financial features.

The results of LR(Fig.~\ref{sub3}, Fig.~\ref{sub4}, Fig.~\ref{sub5}), LGB(Fig.~\ref{sub9}, Fig.~\ref{sub2}, Fig.~\ref{sub10}) and LSTM(Fig.~\ref{sub2}, Fig.~\ref{sub16}, Fig.~\ref{sub18}) very clearly show the advantages of hybrid datasets for bankruptcy prediction. Although the results of RF(Fig.~\ref{sub6}, Fig.~\ref{sub6}, Fig.~\ref{sub8}) and the results of MLP(Fig.~\ref{sub11}, Fig.~\ref{sub12}, Fig.~\ref{sub13}), we can still find the advantage of hybrid datasets, but not very obvious. The results of CNN-1D(Fig.~\ref{sub14}, Fig.~\ref{sub15}, Fig.~\ref{sub16}) are inconclusive, as the performance of financial-related features is comparable to random guessing.
Additionally, the performance of models improves with longer training data periods. This suggests that using a larger data set can capture more accurate trends and patterns that indicate potential bankruptcy. Furthermore, it is worth noting that machine learning models such as LR, RF, and LGB outperform deep learning models such as MLP, CNN-1D, and LSTM.
Overall, hybrid datasets offer significant advantages over single-source datasets for predicting bankruptcy. LightGBM model outperforms all other models in 3 time periods.

\begin{figure}
  \centering
  \begin{subfigure}[b]{0.45\linewidth}
    \includegraphics[width=\linewidth,height=5cm]{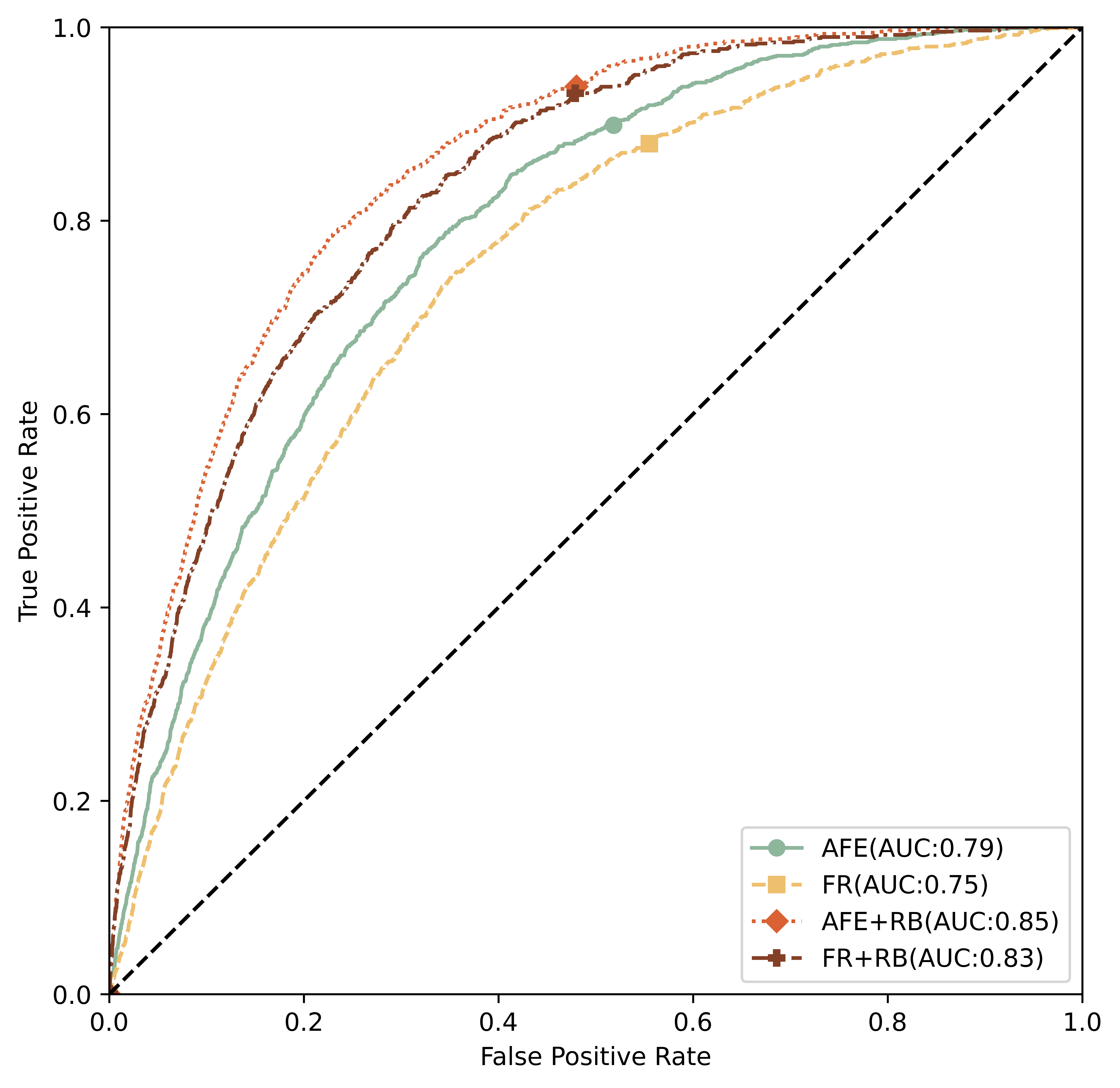}
    \caption{\scriptsize{LGB on 1-year dataset}}
    \label{sub1}
  \end{subfigure}
  \begin{subfigure}[b]{0.45\linewidth}
    \includegraphics[width=\linewidth,height=5cm]{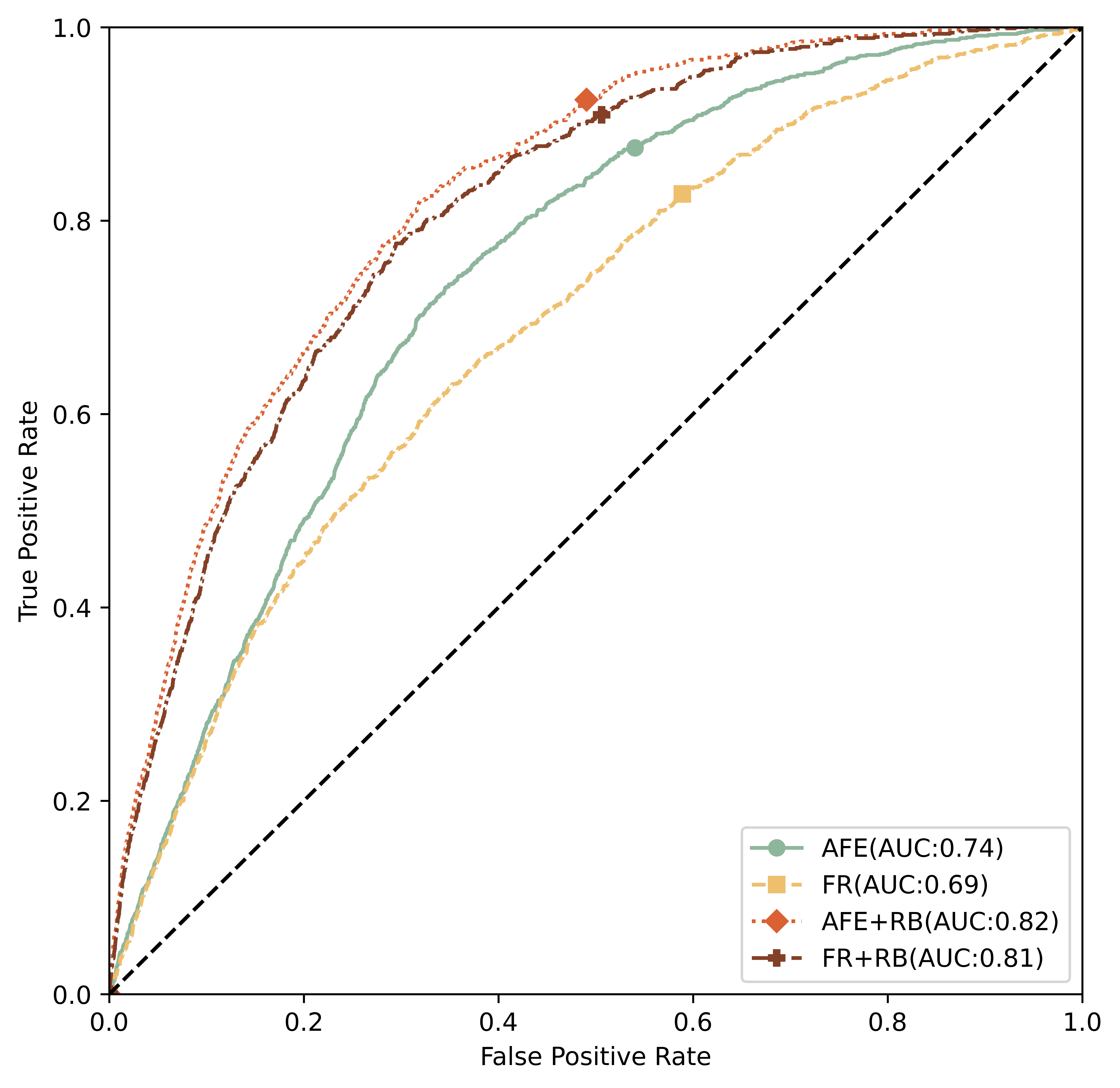}
    \caption{\scriptsize{LSTM on 1-year dataset}}
    \label{sub2}
  \end{subfigure}
  \caption{ROC curve of lightGBM and LSTM on 1-year datasets}
  \label{big}
\end{figure}

\subsection{Performance about Covid period}

As described in Section III, the bankruptcy rate decreases significantly since 2019. It only has a 1\% bankruptcy rate in 2019 and less than a 5\% bankruptcy rate for 2020 and 2021. There are several reasons for the drop in the bankruptcy rate. First, the implementation of fiscal stimulus policies. Many countries adopted large-scale fiscal stimulus policies to ease the economic pressure caused by the epidemic, such as providing loans, tax cuts, and direct funding to businesses. Implementing these policies may help companies maintain cash flow and reduce the risk of bankruptcy. Second, debt moratorium and grace period. Many companies obtained debt moratorium and grace period arrangements during the pandemic, which allowed them to delay debt repayment, thereby easing short-term financial stress and reducing the risk of bankruptcy.

However, this downward trend in bankruptcy rates may only be temporary, as these policies and arrangements may be unsustainable and companies are facing various uncertainties and challenges. For now, we can not see any evidence directly from the data but just observe that the distribution of both pre-Covid set and post-Covid set drift a lot from the training set. The experimental results verify this observation table \ref{all}. 

More than 50\% results show that the model performances of pre-Covid sets and post-Covid sets are better than those of testing sets, which is contrary to common sense. Furthermore, we find the hybrid datasets perform less favorable for both pre-Covid and post-Covid time period. We assume that this is because the reporting behavior of companies changed during the pandemic period, which means companies may not submit or report their restructuring behavior in time due to the pandemic. This inconsistency on reported behavior data will confuse the model thus make the prediction performance not as good as testing set.

\begin{table*}[htbp]
\vspace{-0.8cm}
\scriptsize
    \centering
    \caption{AUC of models on testing, pre-Covid and post-Covid sets}
    \begin{adjustbox}{center}
    \begin{tabular}{|l|lll|lll|lll|}
    \hline
        \multirow{2}{*}{}&\multicolumn{3}{|c|}{\textbf{1-year}} & \multicolumn{3}{|c|}{\textbf{2-year}}& \multicolumn{3}{|c|}{\textbf{3-year}} \\ \cline{2-10}
         & \textbf{Testing} & \textbf{Pre-Covid} & \textbf{Post-Covid} & \textbf{Testing} & \textbf{Pre-Covid} & \textbf{Post-Covid} & \textbf{Testing} & \textbf{Pre-Covid} & \textbf{Post-Covid} \\ \hline
         \multicolumn{10}{|c|}{\textbf{LR}}\\ \hline
        \textbf{AFE} & 0.7522 & 0.7539 & \textbf{0.7669} & 0.7494 & 0.7486 & \textbf{0.7632} & \textbf{0.7702} & 0.7599 & 0.7548 \\ 
        \textbf{FR} & 0.7231 & \textbf{0.7493} & 0.7425 & 0.7384 & \textbf{0.7621} & 0.7605 & 0.7706 & \textbf{0.7755} & 0.7660 \\ 
        \textbf{AFE+RB} & \textbf{0.8234} & 0.7433 & 0.7129 & \textbf{0.8556} & 0.7636 & 0.7124 & \textbf{0.8767} & 0.7755 & 0.7226 \\
        \textbf{FR+RB} & \textbf{0.8118} & 0.7375 & 0.6948 & \textbf{0.8642} & 0.7758 & 0.7037 & \textbf{0.8918} & 0.7885 & 0.7178 \\ \hline
         \multicolumn{10}{|c|}{\textbf{RF}}\\ \hline
        \textbf{AFE } & 0.7894 & \textbf{0.8018} & 0.7786 & 0.7860 & \textbf{0.8077} & 0.7775 & 0.8024 & 0.8052 & \textbf{0.8078} \\
        \textbf{FR} & 0.7564 & \textbf{0.7687} & 0.7337 & 0.7676 & \textbf{0.7838} & 0.7454 & 0.7836 & 0.7810 & \textbf{0.7873} \\
        \textbf{AFE+RB} & 0.7733 & 0.7739 & \textbf{0.8195} & 0.7733 & 0.7658 & \textbf{0.8423} & 0.7876 & 0.7711 & \textbf{0.8589} \\ 
        \textbf{FR+RB} & 0.7294 & 0.7316 & \textbf{0.7941} & 0.745 & 0.7309 & \textbf{0.8164} & 0.7524 & 0.7286 & \textbf{0.8418} \\ \hline
         \multicolumn{10}{|c|}{\textbf{LGB}}\\ \hline
        \textbf{AFE } & 0.7887 & 0.7980 & \textbf{0.8092} & 0.7925 & 0.7976 & \textbf{0.8156} & 0.8133 & 0.8108 & \textbf{0.8147} \\
        \textbf{FR} & 0.7490 & 0.7629 & \textbf{0.7741} & 0.7634 & 0.7752 & \textbf{0.7912} & \textbf{0.7990} & 0.7865 & 0.7903 \\
        \textbf{AFE+RB} & \textbf{0.8542} & 0.7732 & 0.7705 & \textbf{0.8783} & 0.8065 & 0.7623 & \textbf{0.8930} & 0.8168 & 0.7621 \\ 
        \textbf{FR+RB} & \textbf{0.8312} & 0.7421 & 0.7226 & \textbf{0.8706} & 0.7759 & 0.7299 & \textbf{0.8903} & 0.7919 & 0.7313 \\ \hline  
         \multicolumn{10}{|c|}{\textbf{MLP}}\\ \hline
        \textbf{AFE } & 0.7408 & 0.7383 & \textbf{0.7536} & 0.7428 & 0.7447 & \textbf{0.7752} & \textbf{0.7489} & 0.7240 & 0.7263 \\
        \textbf{FR} & 0.7282 & \textbf{0.7482} & 0.7299 & 0.7204 & \textbf{0.7585} & 0.7537 & 0.7301 & \textbf{0.7462} & 0.7329 \\ 
        \textbf{AFE+RB} & \textbf{0.8109} & 0.6799 & 0.6775 & \textbf{0.8014} & 0.6841 & 0.7032 & \textbf{0.7743} & 0.6710 & 0.6643 \\ 
        \textbf{FR+RB} & \textbf{0.7980} & 0.6904 & 0.674 & \textbf{0.8145} & 0.6946 & 0.6775 & \textbf{0.8218} & 0.7238 & 0.6902 \\ \hline
         \multicolumn{10}{|c|}{\textbf{CNN-1D}}\\ \hline
        \textbf{AFE } & 0.7277 & 0.7288 & \textbf{0.7335} & 0.5607 & 0.5857 & \textbf{0.6705} & 0.4318 & 0.4629 & \textbf{0.5317} \\
        \textbf{FR} & 0.6153 & 0.6495 & \textbf{0.7055} & 0.4922 & 0.4954 & \textbf{0.5524} & 0.7165 & 0.7196 & \textbf{0.7333} \\
        \textbf{AFE+RB} & \textbf{0.7442} & 0.7090 & 0.7188 & \textbf{0.7306} & 0.6361 & 0.6229 & \textbf{0.7243} & 0.6298 & 0.5995 \\ 
        \textbf{FR+RB} & \textbf{0.7508} & 0.6456 & 0.6504 & \textbf{0.7139} & 0.615 & 0.6141 & \textbf{0.7217} & 0.6339 & 0.6114 \\ \hline 
        \multicolumn{10}{|c|}{\textbf{LSTM}}\\ \hline
        \textbf{AFE } & 0.7404 & 0.7468 & \textbf{0.7579} & 0.7066 & 0.7036 & \textbf{0.7406} & 0.7193 & 0.7248 & \textbf{0.7363} \\
        \textbf{FR} & 0.6879 & 0.7247 & \textbf{0.7497} & 0.7064 & 0.7369 & \textbf{0.7628} & 0.7419 & 0.7554 & \textbf{0.7563} \\
        \textbf{AFE+RB} & \textbf{0.8245} & 0.7342 & 0.7145 & \textbf{0.8158} & 0.7094 & 0.6924 & \textbf{0.8046} & 0.7236 & 0.6951 \\ 
        \textbf{FR+RB} & \textbf{0.8087} & 0.7159 & 0.6920 & \textbf{0.8211} & 0.7461 & 0.7146 & \textbf{0.8187} & 0.7470 & 0.7114 \\ \hline
    \end{tabular}
    \label{all}        
    \end{adjustbox}
    \vspace{-0.8cm}
\end{table*}

\section{Conclusion}
In conclusion, this study introduce the historical background of bankruptcy prediction models, which have traditionally used accounting-based ratios as input variables. It also presents a new approach to improving these models by incorporating data on reported corporate restructuring behavior. The study compares six models and identifies the most suitable one for predicting bankruptcy. The study validates the effectiveness of the hybrid dataset and analyzes the potential drift of the model during the pandemic. The experimental results demonstrate that utilizing a hybrid dataset can enhance the performance of bankruptcy prediction models by 4\%-13\% compared to using a single source dataset. We also assess the performance of these models during the pandemic and analyze their drift. This study offers valuable insights into bankruptcy prediction models and highlights areas for future research. The findings of this study can assist SMEs in identifying risks, adapting their business strategies, and enhancing their competitiveness and stability in a timely manner. Furthermore, the proposed bankruptcy prediction model can be used to assign credit ratings for SMEs and provide credit endorsements, thereby facilitating their growth. Finally, this study can assist governments and social organizations in identifying potential financial crises and implementing proactive measures to mitigate adverse economic and social impacts.

\subsubsection{Acknowledgements} We thank the Luxembourg National Research Fund (FNR) under Grant 15403349 and Yoba S.A. for supporting this work.

\bibliographystyle{IEEEtran}
\bibliography{IEEEabrv,reference}

\appendix
\section{Appendix A}

\begin{figure}
  \centering
  \begin{subfigure}[b]{0.45\linewidth}
    \includegraphics[width=\linewidth,height=5cm]{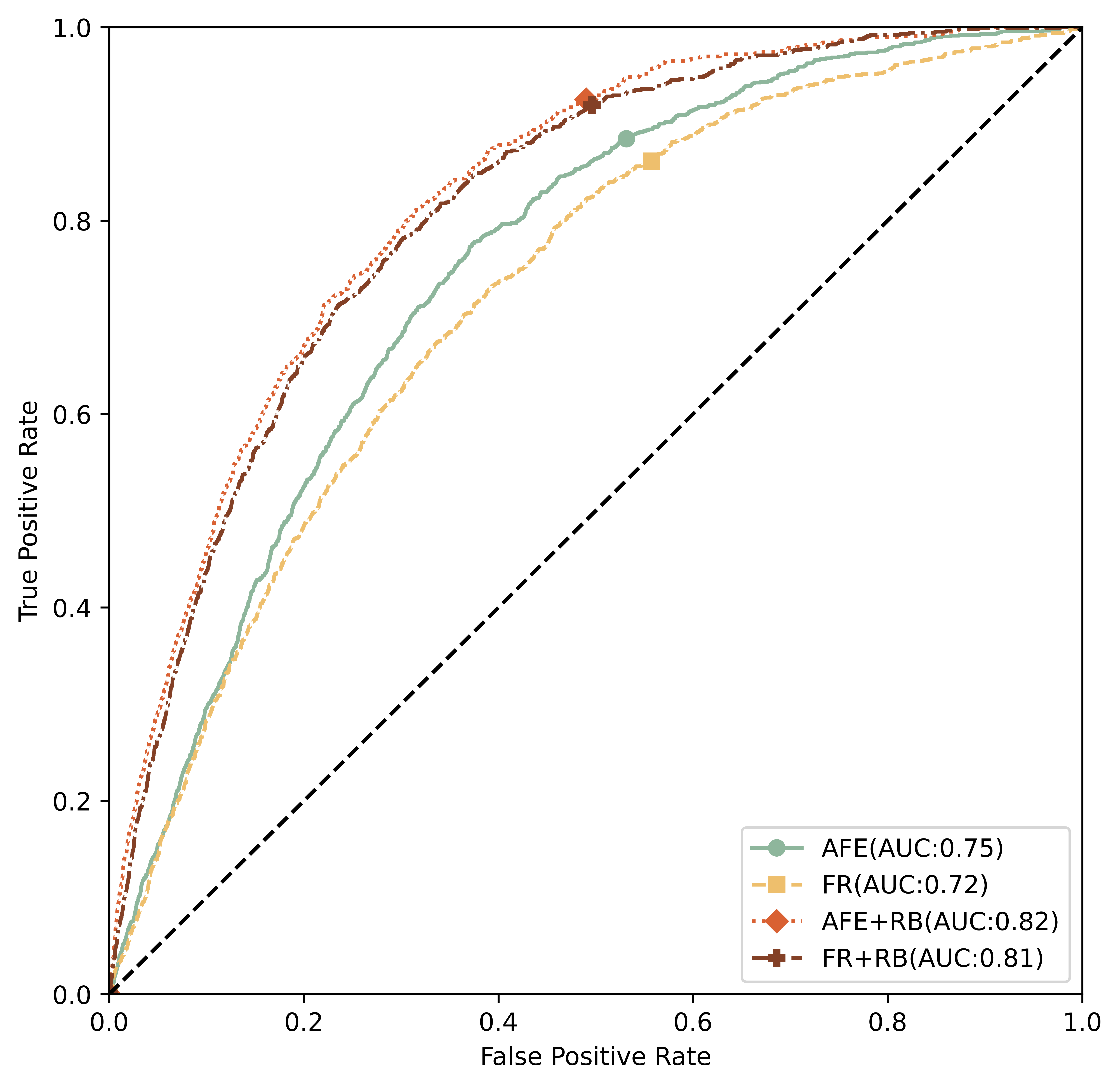}
    \caption{\scriptsize{LR on 1-year dataset}}
    \label{sub3}
  \end{subfigure}
  \begin{subfigure}[b]{0.45\linewidth}
    \includegraphics[width=\linewidth,height=5cm]{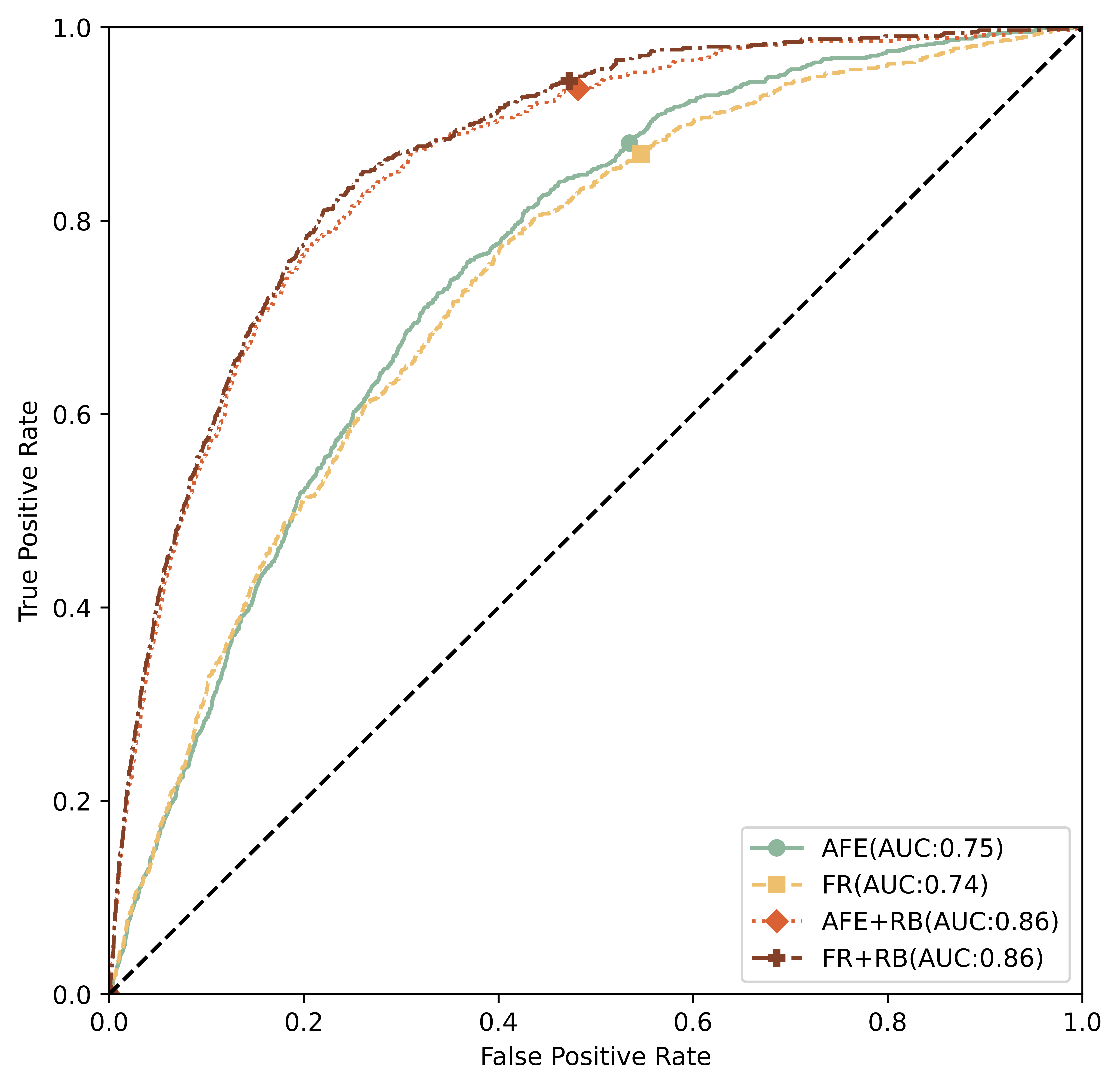}
    \caption{\scriptsize{LR on 2-year dataset}}
    \label{sub4}
  \end{subfigure}
\medskip

\begin{subfigure}[b]{0.45\linewidth}
\includegraphics[width=\linewidth,height=5cm]{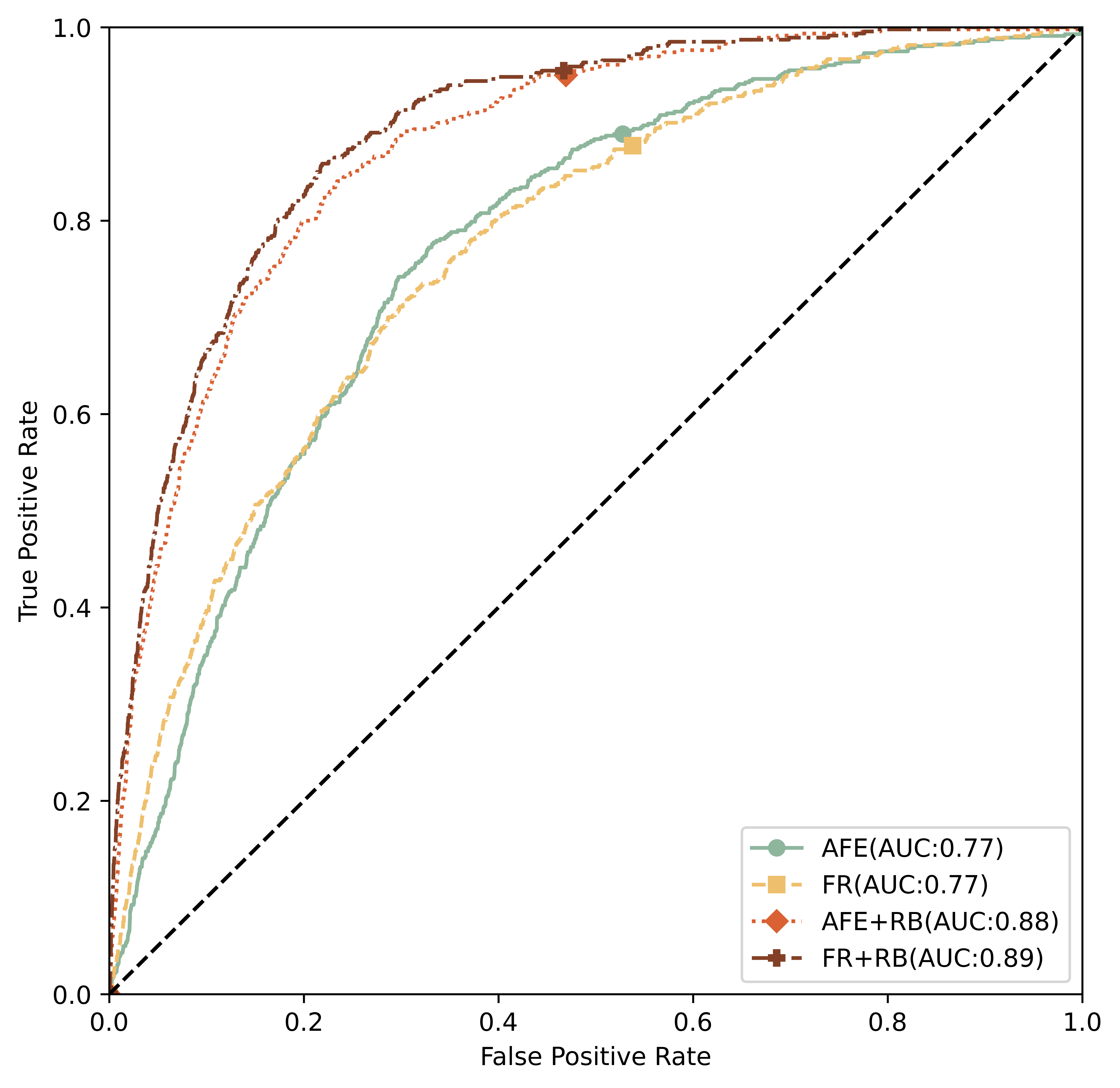}
\caption{\scriptsize{LR on 3-year dataset}}
\label{sub5}
\end{subfigure}
\begin{subfigure}[b]{0.45\linewidth}
\includegraphics[width=\linewidth,height=5cm]{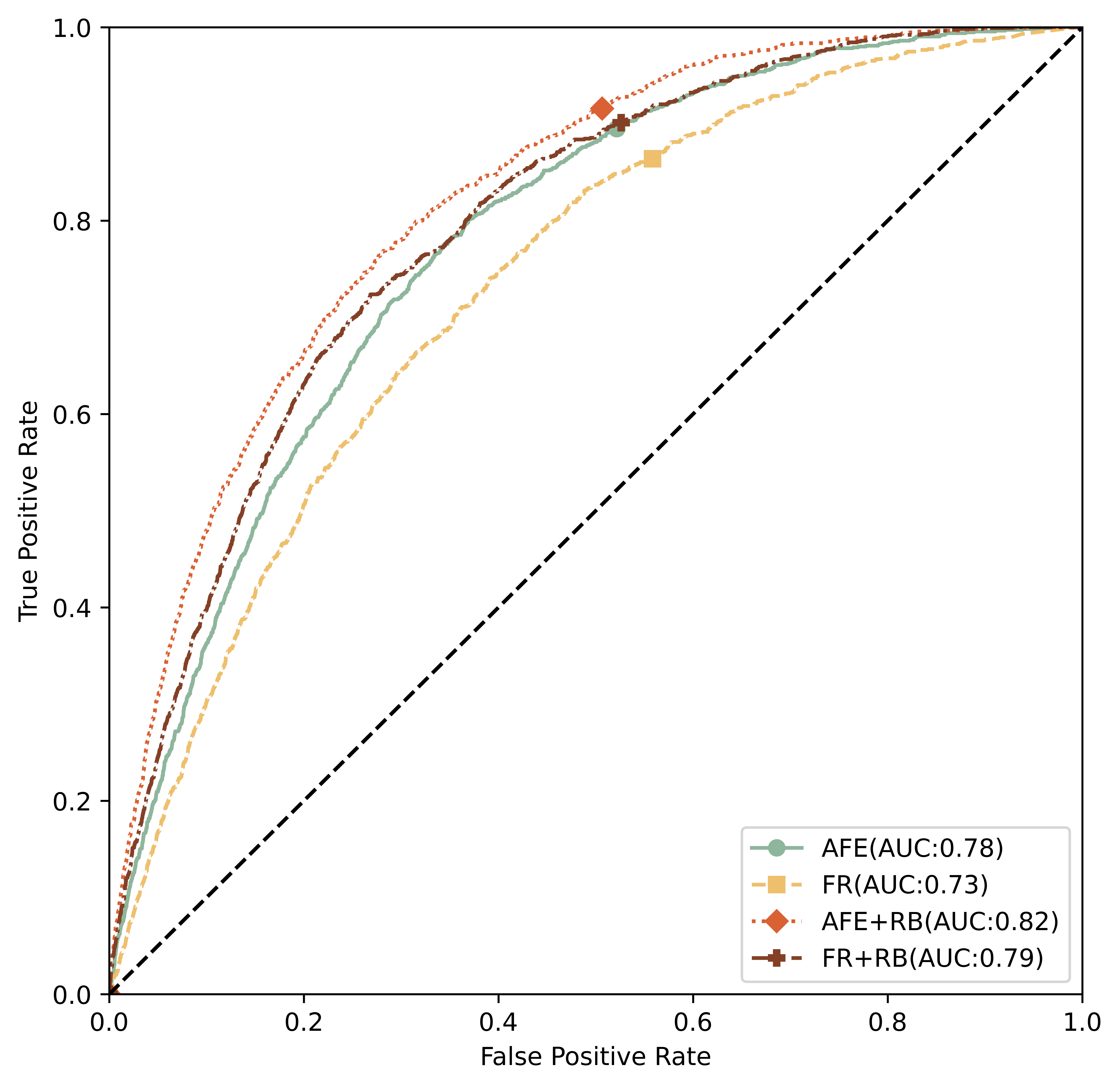}
\caption{\scriptsize{RF on 1-year dataset}}
\label{sub6}
\end{subfigure}
\medskip

\begin{subfigure}[b]{0.45\linewidth}
\includegraphics[width=\linewidth,height=5cm]{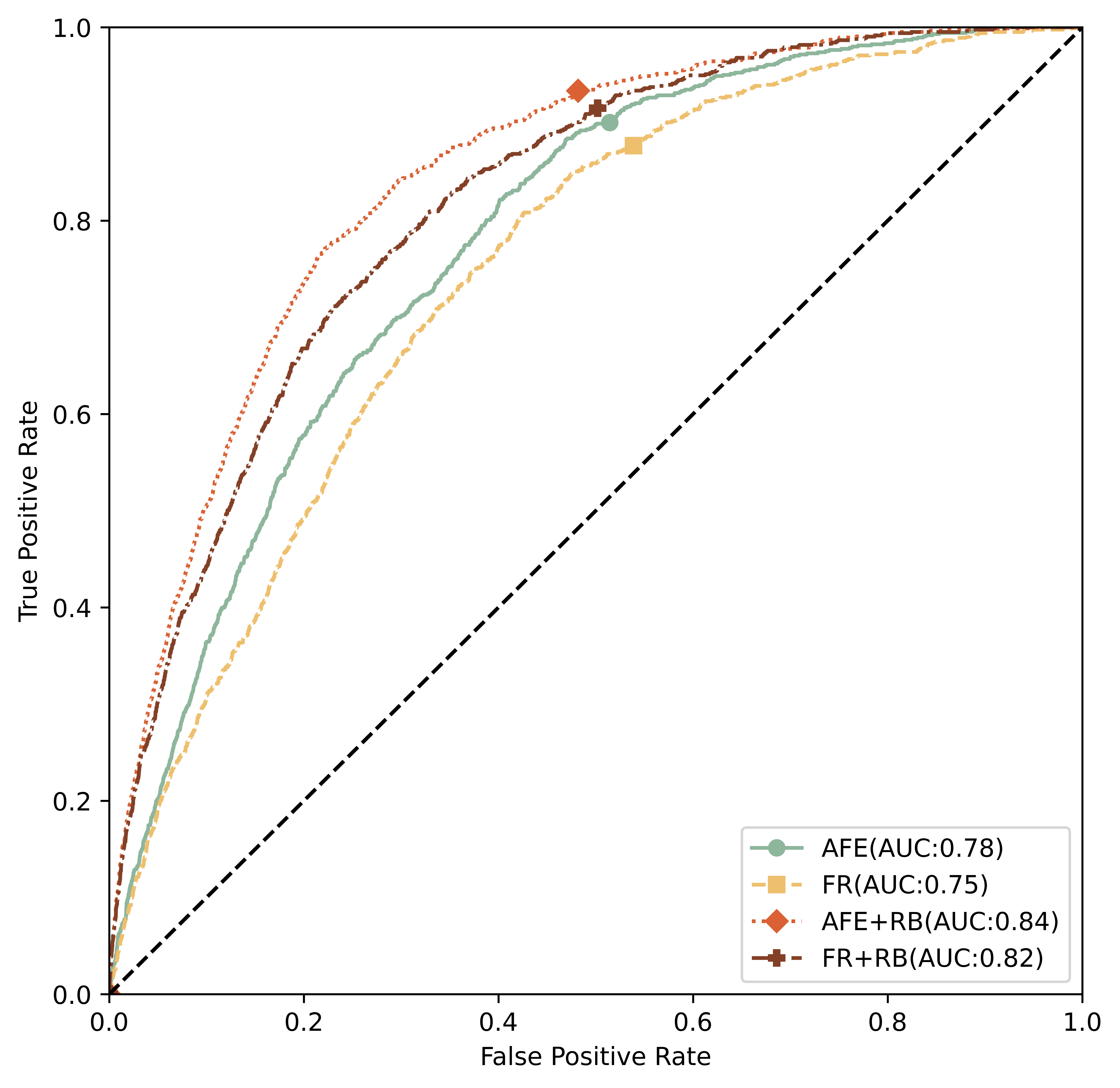}
\caption{\scriptsize{RF on 2-year dataset}}
\label{sub7}
\end{subfigure}
\begin{subfigure}[b]{0.45\linewidth}
\includegraphics[width=\linewidth,height=5cm]{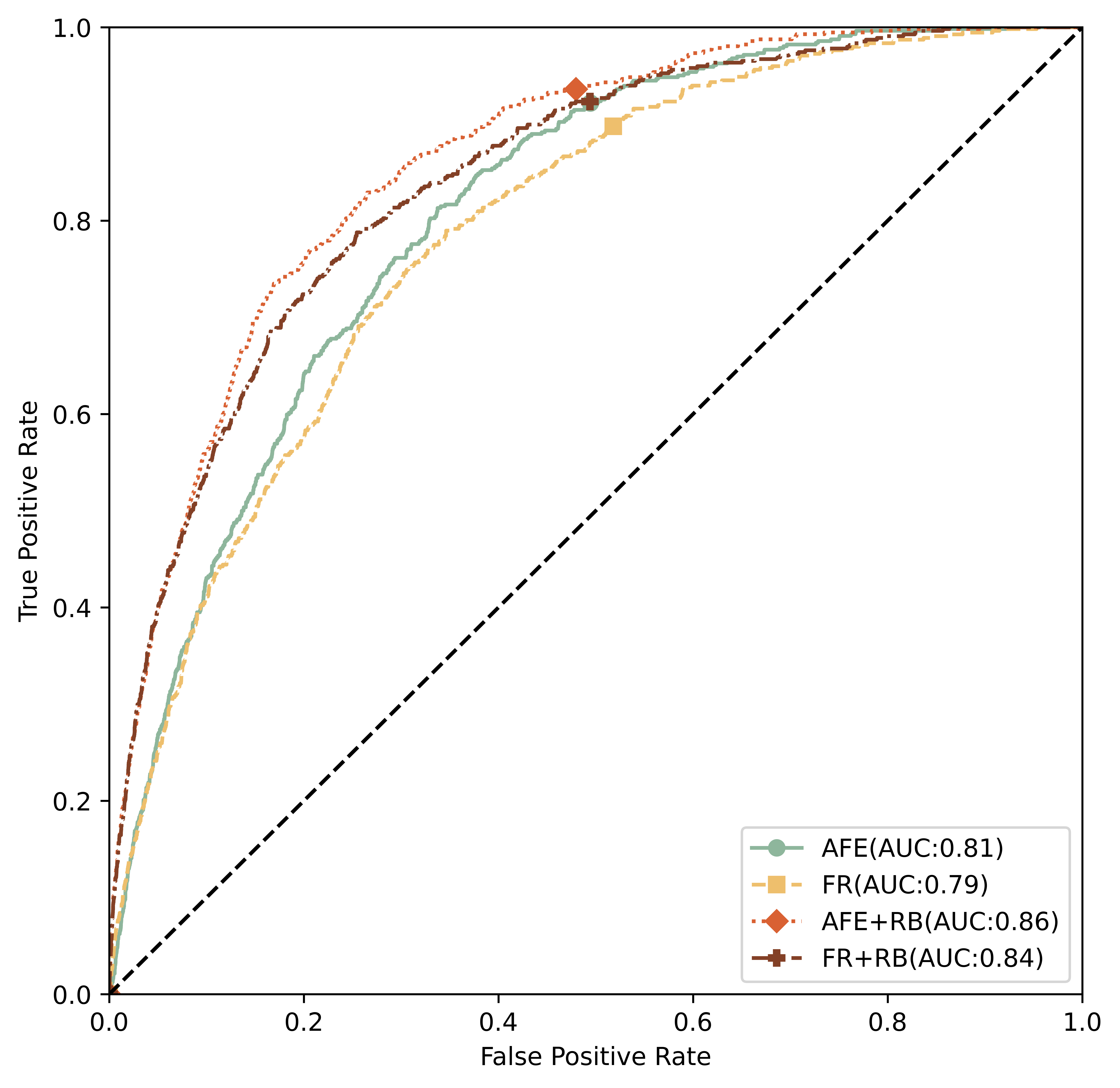}
\caption{\scriptsize{RF on 3-year dataset}}
\label{sub8}
\end{subfigure}
\medskip

\begin{subfigure}[b]{0.45\linewidth}
\includegraphics[width=\linewidth,height=5cm]{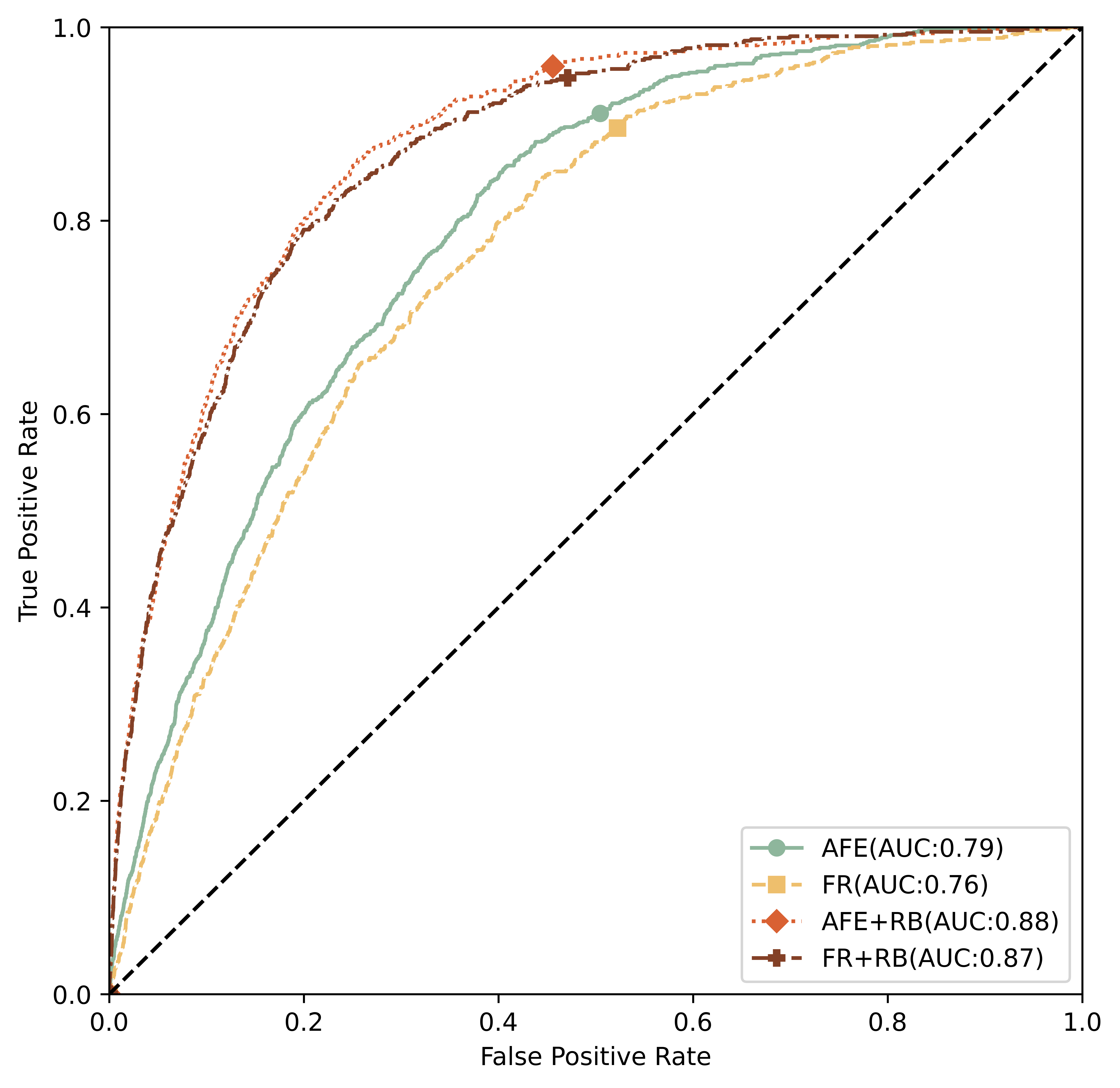}
\caption{\scriptsize{LGB on 2-year dataset}}
\label{sub9}
\end{subfigure}
\begin{subfigure}[b]{0.45\linewidth}
\includegraphics[width=\linewidth,height=5cm]{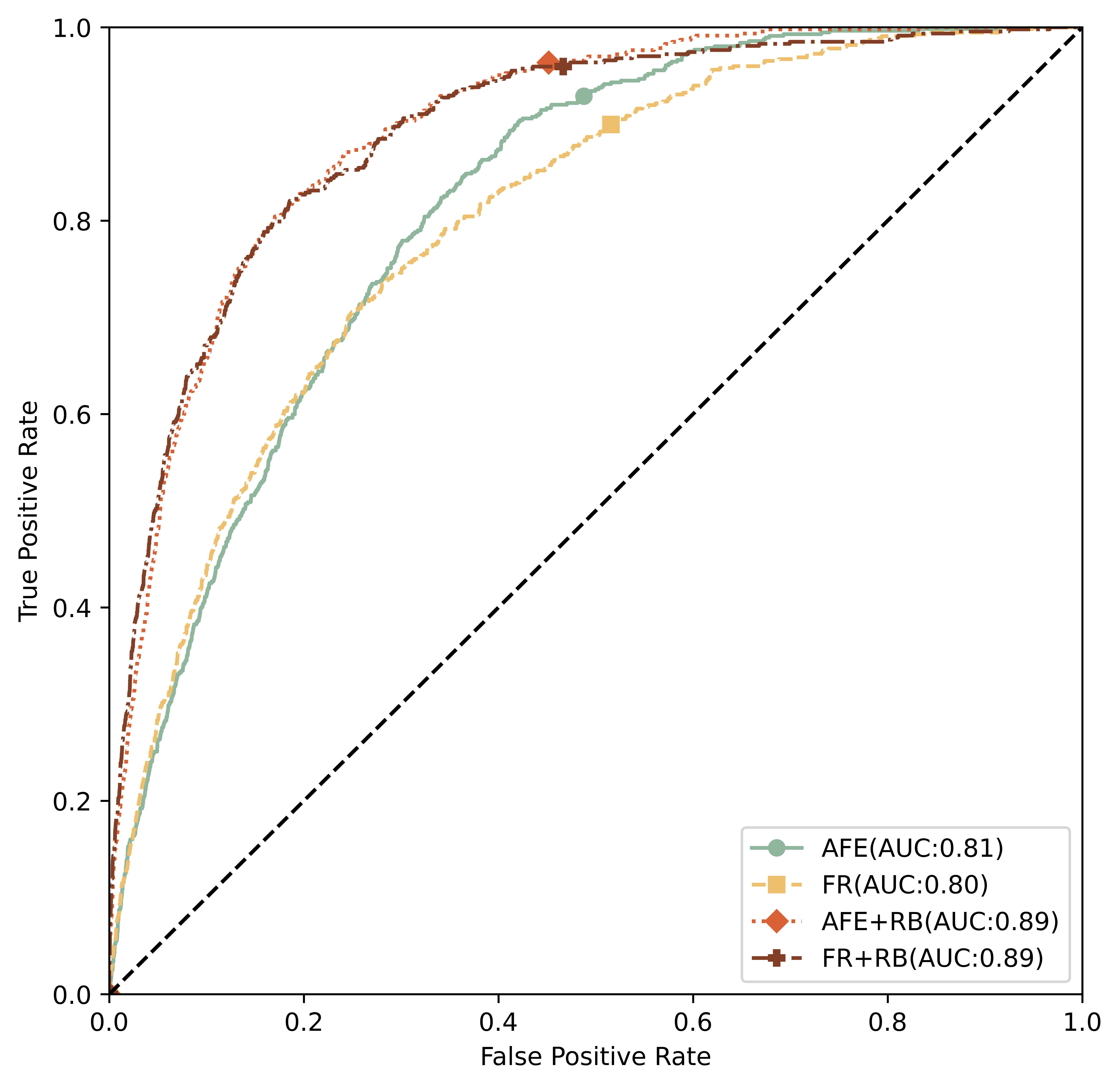}
\caption{\scriptsize{LGB on 3-year dataset}}
\label{sub10}
\end{subfigure}
\medskip
\end{figure}

\begin{figure}
\ContinuedFloat
\begin{subfigure}[b]{0.45\linewidth}
\includegraphics[width=\linewidth,height=5cm]{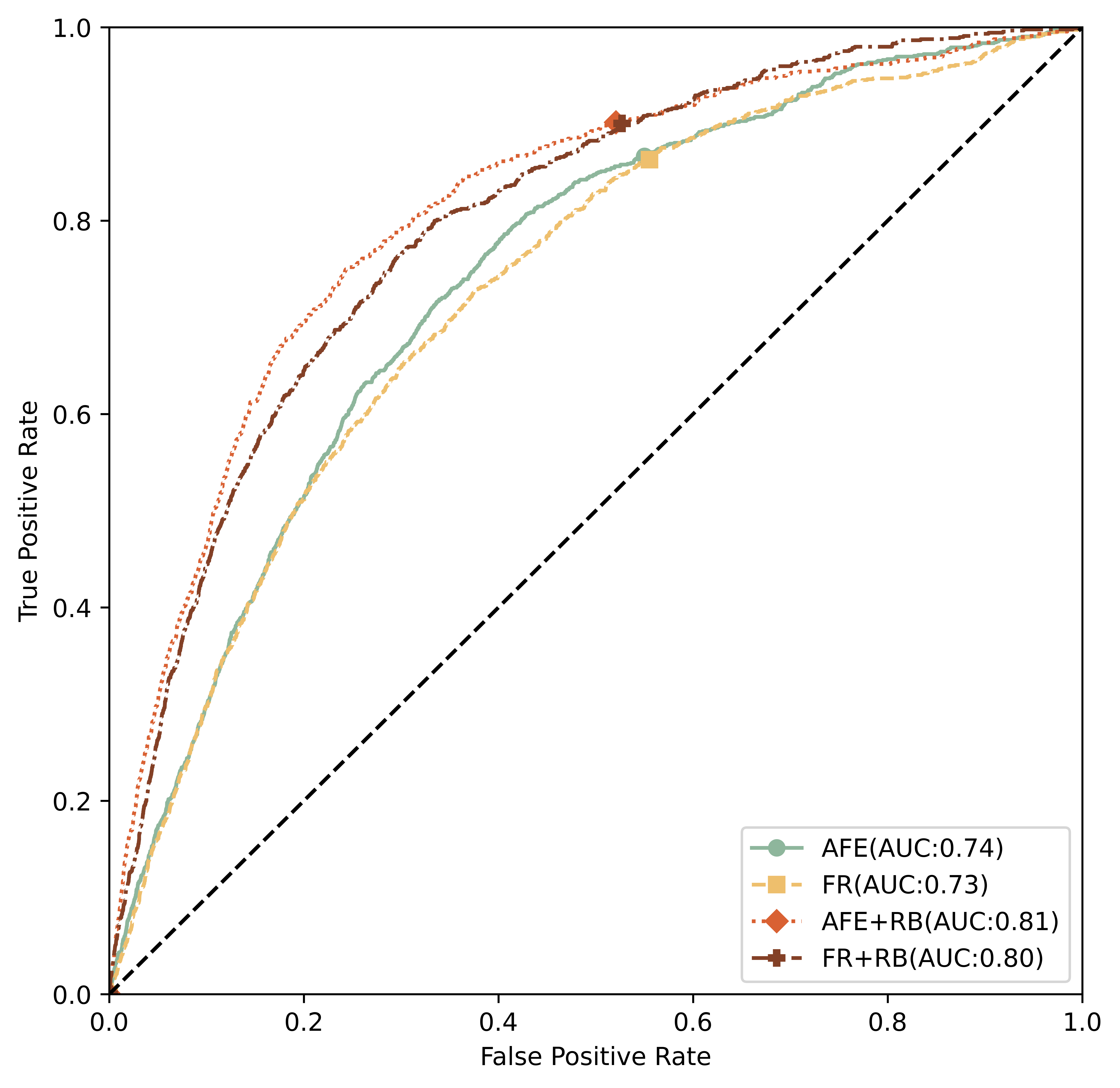}
\caption{\scriptsize{MLP on 1-year dataset}}
\label{sub11}
\end{subfigure}
\begin{subfigure}[b]{0.45\linewidth}
\includegraphics[width=\linewidth,height=5cm]{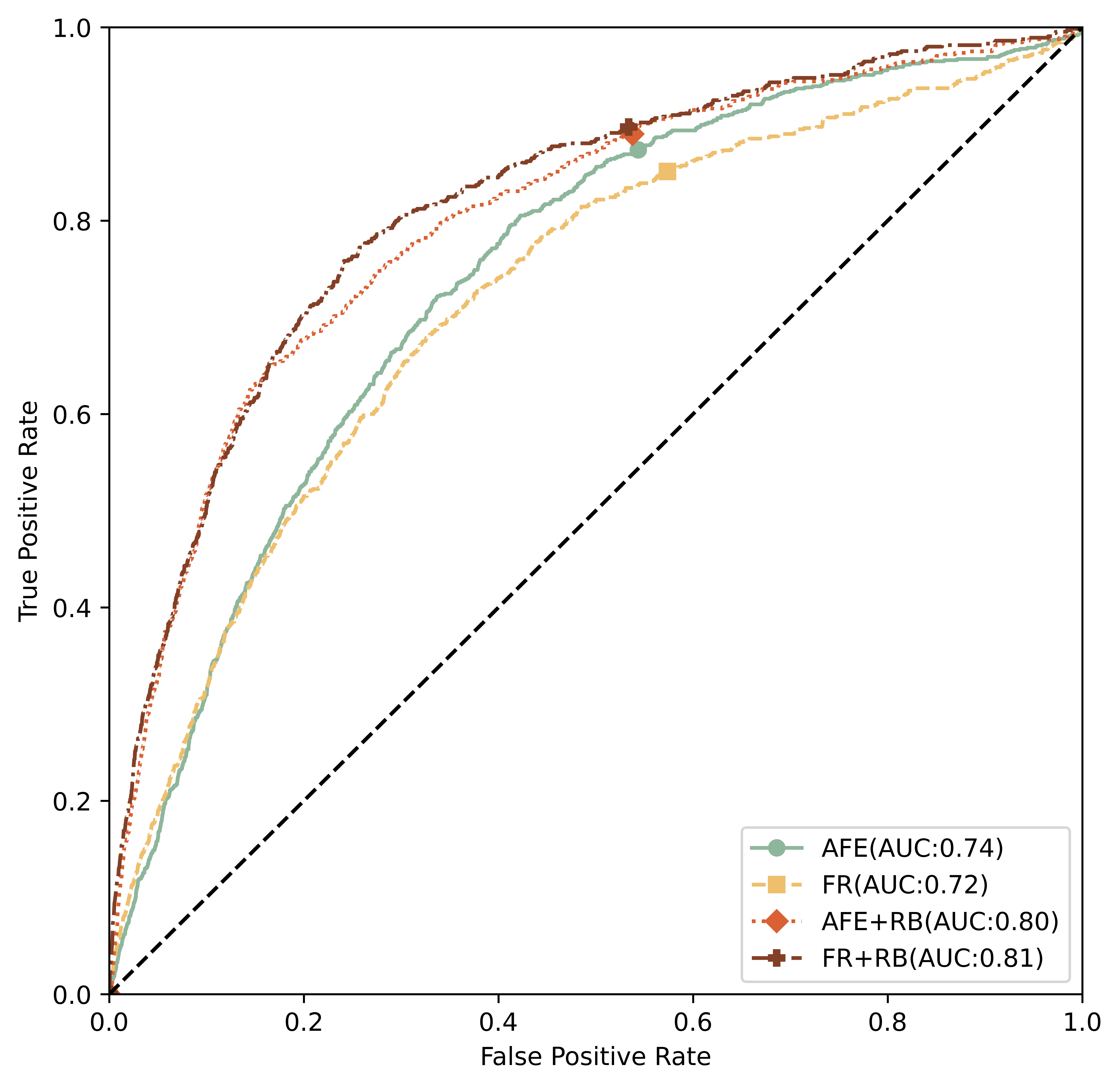}
\caption{\scriptsize{MLP on 2-year dataset}}
\label{sub12}
\end{subfigure}
\medskip

\begin{subfigure}[b]{0.45\linewidth}
\includegraphics[width=\linewidth,height=5cm]{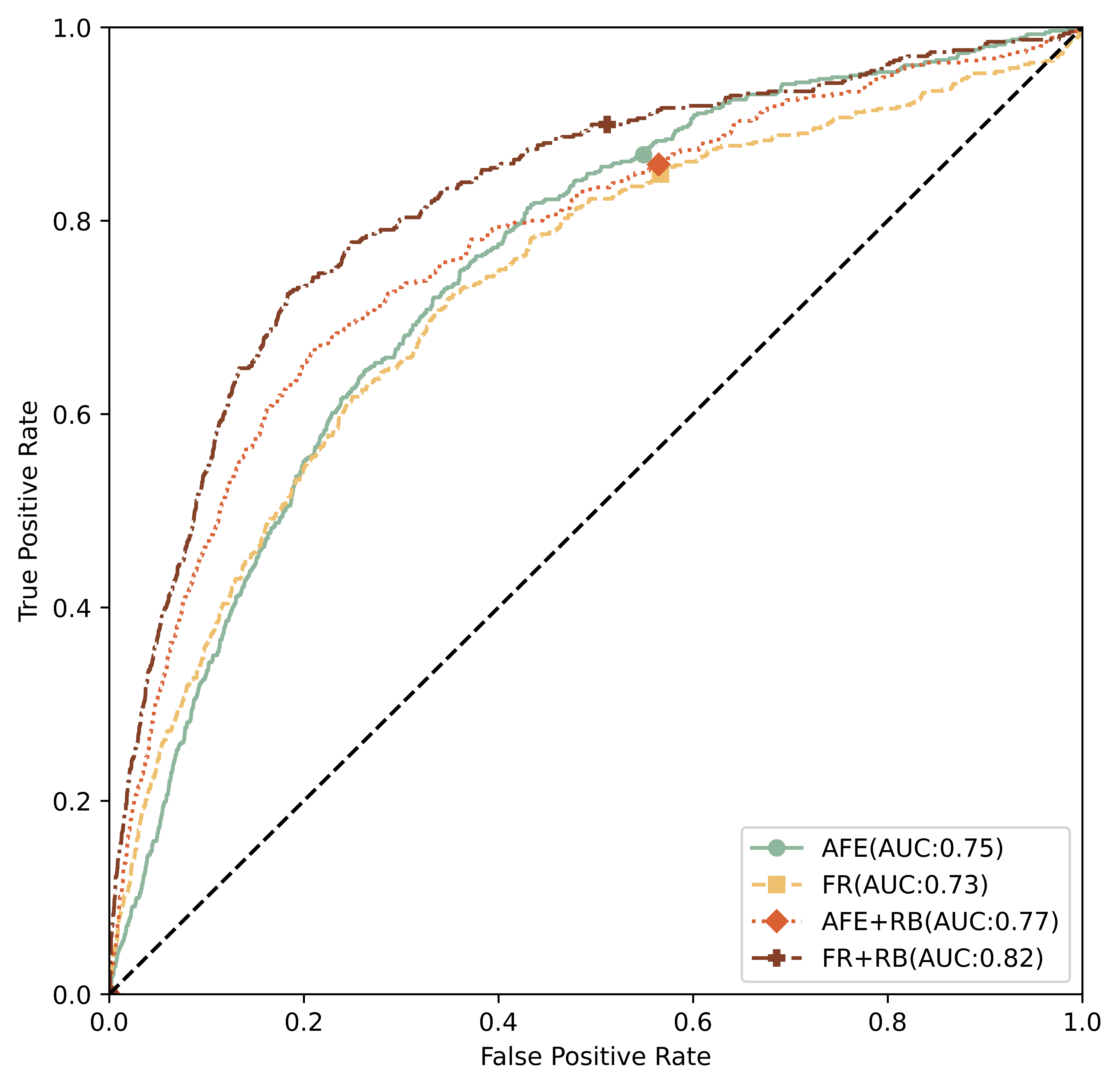}
\caption{\scriptsize{MLP on 3-year dataset}}
\label{sub13}
\end{subfigure}
\begin{subfigure}[b]{0.45\linewidth}
\includegraphics[width=\linewidth,height=5cm]{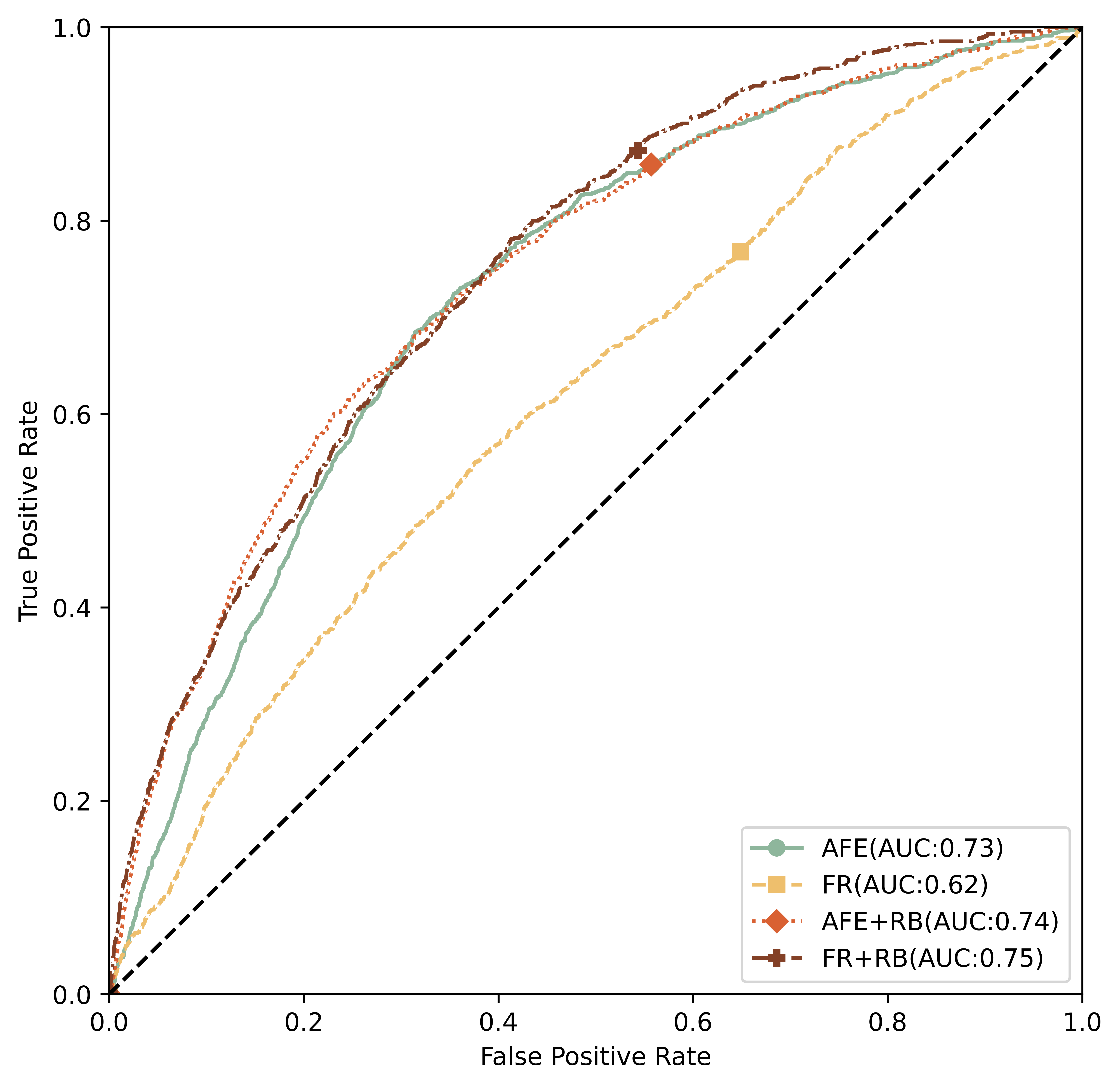}
\caption{\scriptsize{CNN-1D on 1-year dataset}}
\label{sub14}
\end{subfigure}
\medskip

\begin{subfigure}[b]{0.45\linewidth}
\includegraphics[width=\linewidth,height=5cm]{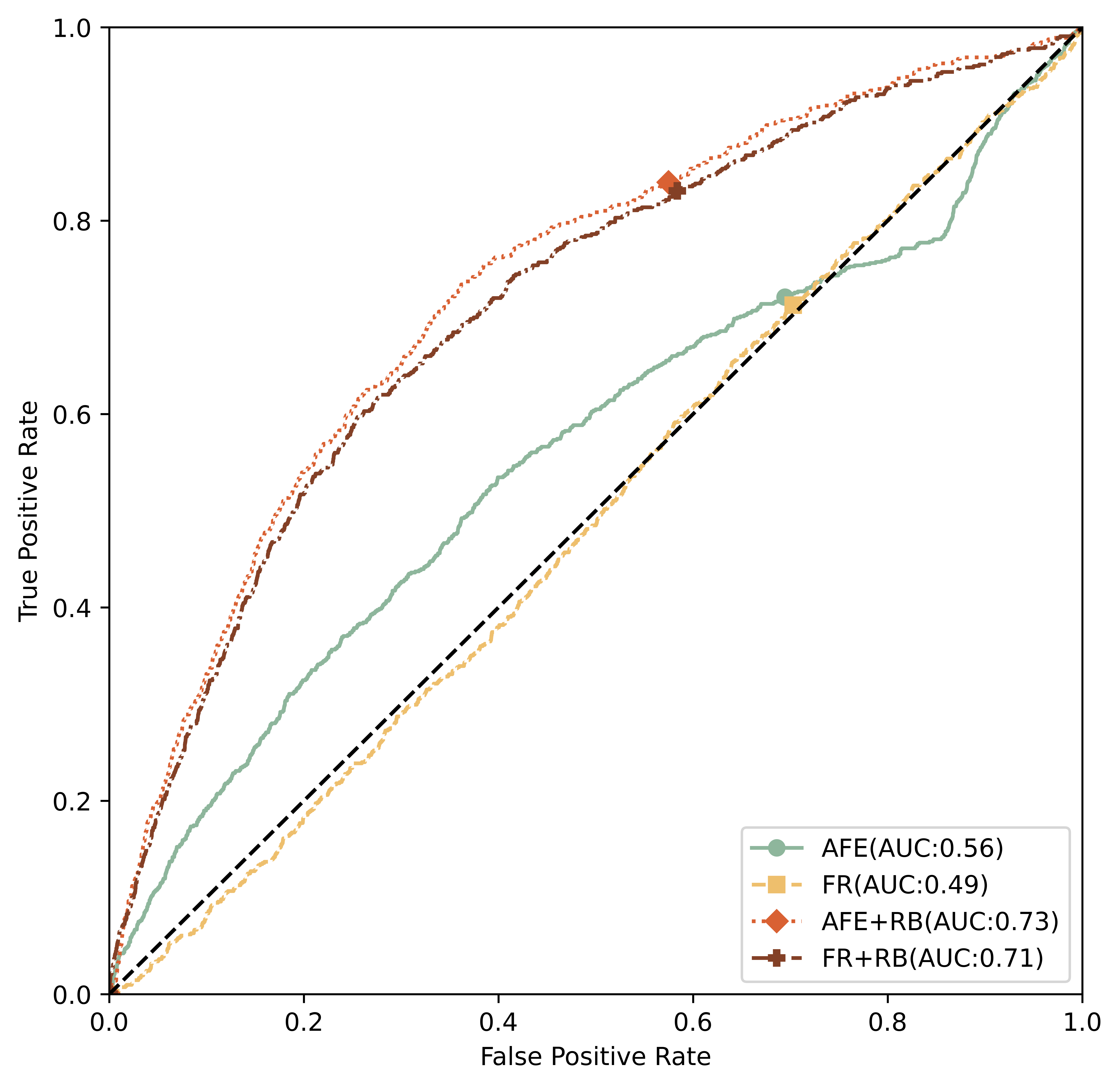}
\caption{\scriptsize{CNN-1D on 2-year dataset}}
\label{sub15}
\end{subfigure}
\begin{subfigure}[b]{0.45\linewidth}
\includegraphics[width=\linewidth,height=5cm]{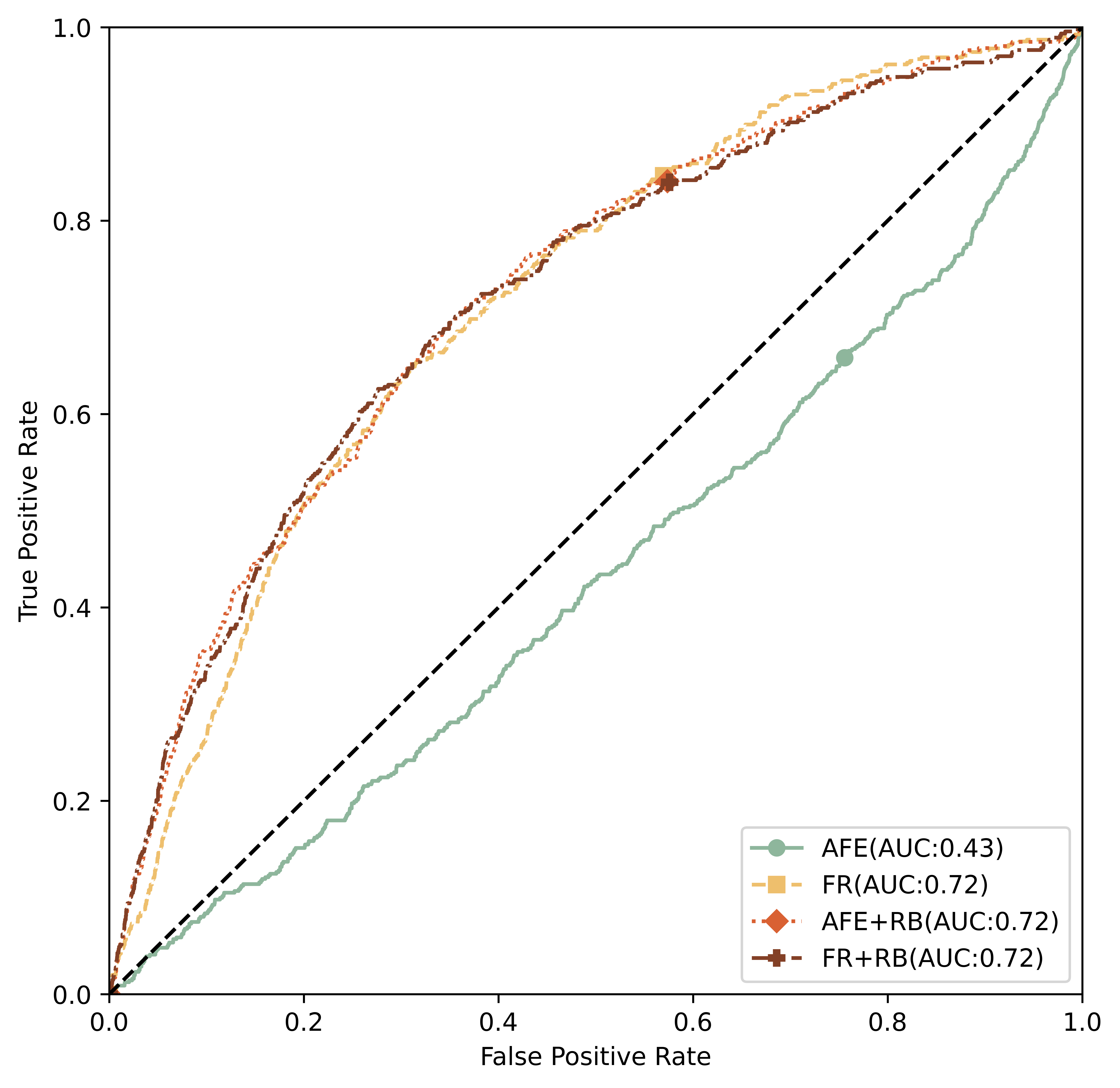}
\caption{\scriptsize{CNN-1D on 3-year dataset}}
\label{sub16}
\end{subfigure}
\medskip

\begin{subfigure}[b]{0.45\linewidth}
\includegraphics[width=\linewidth,height=5cm]{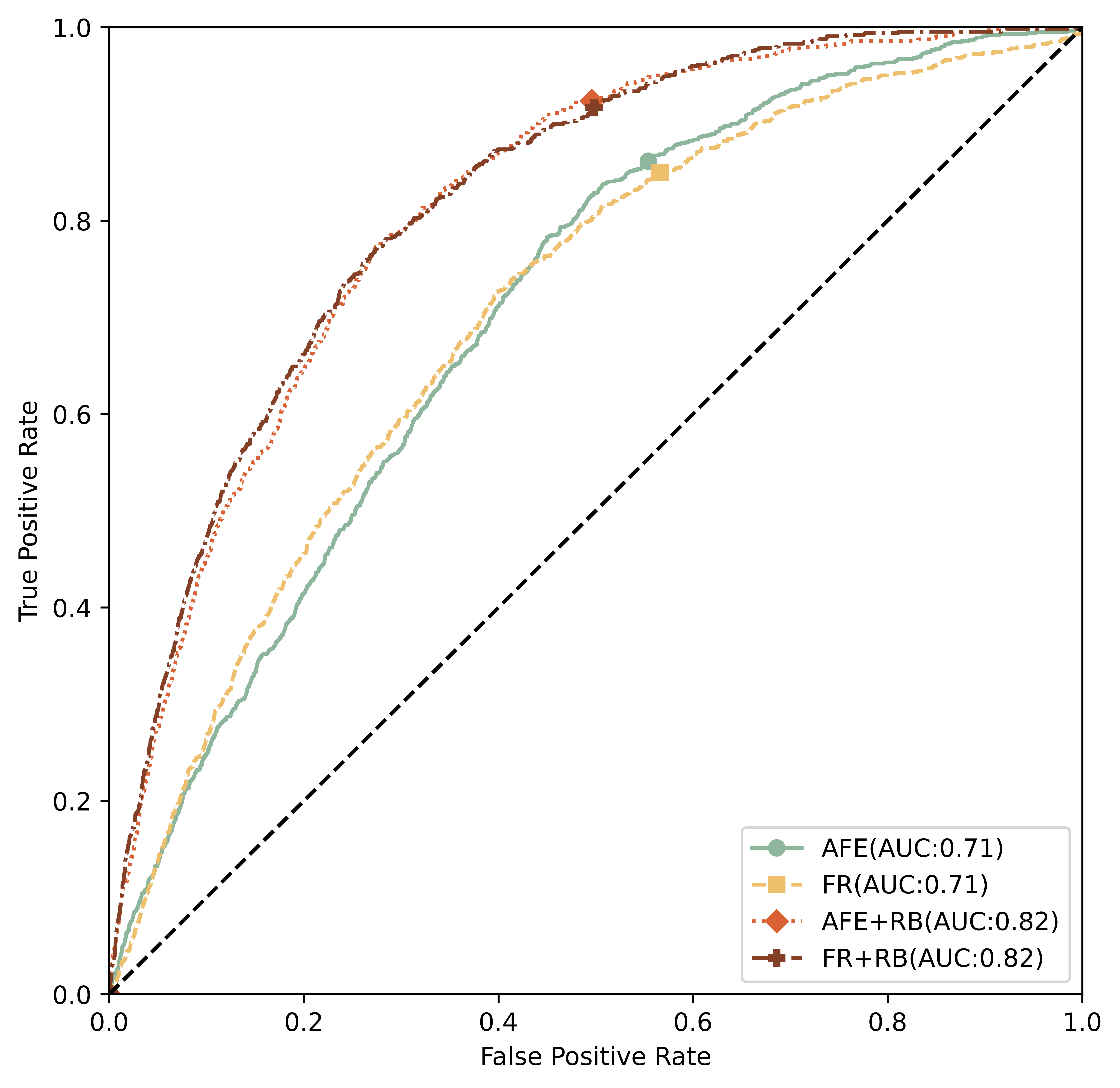}
\caption{\scriptsize{LSTM on 2-year dataset}}
\label{sub17}
\end{subfigure}
\begin{subfigure}[b]{0.45\linewidth}
\includegraphics[width=\linewidth,height=5cm]{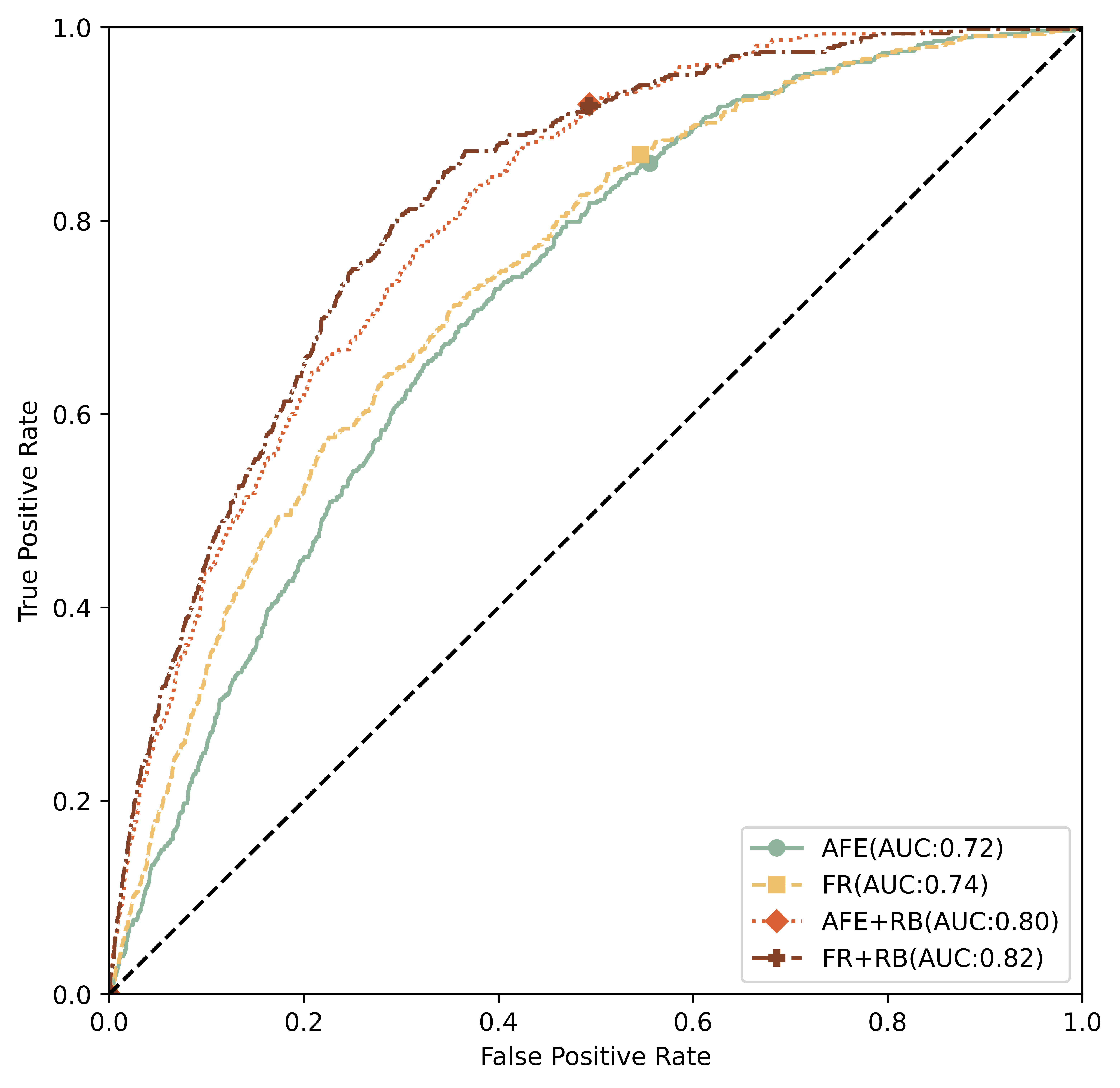}
\caption{\scriptsize{LSTM on 3-year dataset}}
\label{sub18}
\end{subfigure}
\caption{The rest results for ROC curve of 6 models on 3 datasets}
\label{rest}
\end{figure}

\end{document}